# Laser-Induced Current Transients in Ultrafast All-Optical Switching of Metallic Spin Valves

Serban Lepadatu[1,*], Mohammed Gija[1], Alexey Dobrynin[2], Kevin McNeill[2], Mark Gubbins[2], Tim Mercer[1], Steven M. McCann[1], Philip Bissell[1]

[1]*Jeremiah Horrocks Institute for Mathematics, Physics and Astronomy, University of Lancashire, Preston PR1 2HE, U.K.*

[2]*Seagate Technology, 1 Disc Drive, Derry, BT48 0BF, U.K.*

## Abstract

All-optical switching in a ferromagnetic spin valve is studied here using atomistic spin drift-diffusion dynamics, which includes contributions from spin pumping and superdiffusive transport. We find the switching is governed principally by spin-polarized currents due to non-equilibrium hot electrons excited by the laser pulse and re-equilibration currents. In particular, an initial superdiffusive forward flow of electrons, polarized by the free layer, is generated. This drives parallel to antiparallel switching of the free layer through accumulation of minority spins at the reference layer. A diffusive backward flow of electrons, repolarized by the reference layer, follows the initial superdiffusive flow as the charge distribution re-equilibrates. Due to the pulse width-dependent asymmetric amplitudes of the forward and backward transients, the latter can drive antiparallel to parallel switching, and create multi-domain structures at higher laser fluences and longer pulses. The results obtained here are in agreement with experimental observations, providing a framework for self-consistent modelling of all-optical switching in metallic heterostructures.

* SLepadatu@lancashire.ac.uk

## Introduction

All-optical switching (AOS) using single-shot ultrafast laser pulses has been demonstrated experimentally in a range of materials and structures. Ferrimagnetic GdFeCo alloys have been investigated [1-4], where ultrafast magnetization reversal is governed by the transfer of angular momentum between the antiferromagnetically coupled Gd, Fe, and Co sub-lattices. This mechanism was demonstrated directly in CoGd and CoTb alloys [5]. Single-shot AOS was also observed in a number of multi-layered structures, where the switching depends on the layering structure and material composition. These include synthetic ferrimagnetic stacks [6-8], where interfacial exchange scattering is used to explain switching, showing a complex dependence on laser fluence and pulse width, resulting in formation of multi-domain structures at larger fluences, as well as obtaining AOS with picosecond long pulses [7]. Ferromagnetic Co/Pt multilayers, coupled to ferrimagnetic GdFeCo or FeGd, can also be switched, relying on the exchange coupling to the ferrimagnetic layer as well as angular momentum transfer via spin-polarized hot electron transport [9-14]. Antiferromagnetically coupled Tb/Co and Tb/Fe [15,16] show precessional-like AOS, with switching obtained using laser pulses up to 10 ps duration.

Another important case is that of ferromagnetic multilayers, which include in-plane magnetization Ni/Fe [17,18], as well as out-of-plane magnetization Co/Pt or Co/Ni multilayers [19-23], or multilayers coupled to in-plane magnetization CoFeB [24-26]. AOS in such structures is governed by spin-transfer torques (STT) between the various layers, although the exact origin and details are still under debate. Hot electron generation and associated superdiffusive transport is one mechanism. This was identified as a main contributor in Ref. [25], also ruling out the spin-dependent Seebeck effect mechanism as too weak to explain switching. On the other hand, in a similar structure in Ref. [26], it was concluded that spin pumping and spin-dependent Seebeck effects are the main contributions. In more recent work AOS was demonstrated in Co/Pt spin valves [21-23]. In particular, parallel (P) to antiparallel (AP) switching was obtained at lower laser fluences, while higher laser fluences generated AP to P switching as well as multi-domain structures. Moreover, the laser fluence thresholds for the different switching types was found to have a marked dependence on laser pulse width, as well as on the thickness of the non-magnetic Cu spacer layer. While spin pumping during remagnetization is one mechanism of spin current generation, this cannot explain the observed

AP to P switching, as discussed in Ref. [21]; it was suggested that superdiffusive transport of hot electrons is another possible mechanism. Detailed reproduction of such experimental results using modelling remains a challenge. A semi-classical model of superdiffusive spin transport was developed [27,28], highlighting its importance to ultrafast demagnetization as well as spin current generation in metallic multilayers. On the other hand, using a model based on solution of the Boltzmann equation [29], including hot electron transport and spin pumping, it was concluded the spin current is dominated by the spin pumping contribution. Diffusive contributions were also considered, namely diffusive spin-dependent electron transport and diffusive magnon transport [30], as well as an s-d model for laser-induced magnetization dynamics which includes diffusive spin transport [31,32]. A model based on density-function theory showed magnetization toggle control for timescales less than 100 fs in Co/Pt multilayers [33].

In this work we discuss a two-temperature atomistic spin dynamics model, coupled to a spin drift-diffusion transport model which includes contributions from superdiffusive transport of non-equilibrium hot electrons, spin pumping due to demagnetization and remagnetization, as well as the spin-dependent Seebeck effect. The spin transport model obtains spin torques in each magnetic layer, which self-consistently depend on the multilayer structure, material parameters, laser fluence and pulse width as well as the laser beam intensity profile. Through the use of atomistic spin dynamics, calculations are applicable to both the ferromagnetic and paramagnetic phases, and are able to reproduce time and space-dependent details of AOS and formation of multi-domain structures in three-dimensional metallic heterostructures. Furthermore, we explain the laser fluence thresholds obtained in Ref. [21] for P to AP, as well as AP to P switching and formation of multi-domain structures, their pulse width dependence, as well as dependence on the non-magnetic spacer layer, self-consistently. It is shown here the main source of STT is due to laser-induced current transients, originating in non-equilibrium hot electrons which superdiffuse away from the laser-exposed surface. As electrons accumulate at the opposite surface of the multilayer structure, an equally important current transient is due to diffusive electron backward flow as charges re-equilibrate following the laser pulse excitation. While spin pumping and spin-dependent Seebeck effects are also present, we find their contributions to spin torques are relatively small in the ferromagnetic spin valves studied here.

## Atomistic Spin Drift-Diffusion Model

Atomistic spin drift-diffusion (ASDD) dynamics combines a continuous three-dimensional spin drift-diffusion (SDD) model with a discrete atomistic spin dynamics (ASD) model. The charge current density is written as:

$$\mathbf{J}_C = \sigma S_c \nabla T_e - D_e \nabla n_e - D_e \left[ (\nabla T_e / T_e) n_s + \nabla n_s \right] \tag{1}$$

The first term is the Seebeck effect with coefficient $S_c$, conductivity σ, and electron temperature $T_e$. The second term is the diffusion of an excess charge density $n_e$ with coefficient $D_e$. The third term is the contribution of a non-equilibrium electron density $n_s$ [34]. Non-equilibrium hot electrons, locally excited by a laser pulse, exhibit superdiffusive behavior. While concentration gradients drive a diffusive motion, temperature gradients drive a superdiffusive motion, with current density proportional to $n_s$. The latter follows the continuity relation:

$$\frac{\partial n_s}{\partial t} = -\nabla . \mathbf{J}_C^{(S)} - \frac{n_s}{\tau_s} - \frac{e \eta_s Q \lambda}{hc} \tag{2}$$

Here $\mathbf{J}_C^{(S)} = -D_e \left[ (\nabla T_e / T_e) n_s + \nabla n_s \right]$ is the non-equilibrium current density term from Equation (1). Non-equilibrium electrons relax back to lower energy levels with a mean lifetime $\tau_S$, taken here as 200 fs [27,35], and are created by a source term. The latter is due to a laser with power density $Q$, wavelength $\lambda$, and effective electron-photon excitation number $\eta_s$ (effective number of non-equilibrium electrons excited per photon). Here we use $\lambda$ = 630 nm and $\eta_s$ = 1. The excess charge density follows the usual continuity relation:

$$\frac{\partial n_e}{\partial t} = -\nabla . \mathbf{J}_C \tag{3}$$

As non-equilibrium electrons are driven away from concentration and temperature gradients an excess charge density is created at the top, with electrons accumulating at the bottom of the metallic structure. This results in an additional diffusive current which re-equilibrates the charge distribution after the non-equilibrium electron population has relaxed. Thus, following the initial excitation by the laser pulse, displacement of non-equilibrium electrons drives the accumulation of excess charge, which is maintained even after $n_s$ relaxes locally back to zero. The dissipation of excess charge occurs through diffusion, as included in Equation (3). On the other hand, the excess charge does not influence the dynamics of non-equilibrium electrons, which is dependent only on the source and relaxation terms, as well as the non-equilibrium current density in Equation (2).

Since the laser power density is absorbed based on the Beer-Lambert law, $Q(z) = Q_0 \exp(-z/d)$ where $d$ is the optical absorption length, a significant simplification of Equation (1) may be obtained by noting the electron temperature also decays exponentially away from the laser-exposed surface. Thus, $\nabla T_e / T_e = -\hat{\mathbf{z}}/d$ and the term $v_e = D_e / d$ becomes the mean superdiffusive electron velocity. This is of the order $10^5$ – $10^6$ m/s, as obtained from semi-classical modelling of superdiffusive transport [28], in agreement with experimental observations [35], and directed away from the exposed surface. Boundary conditions are obtained by setting the current density normal to a boundary to zero, giving the non-homogeneous Neumann condition $\nabla n_e . \hat{\mathbf{n}} = \sigma (S_c / D_e) \nabla T_e . \hat{\mathbf{n}} + n_s \mathbf{v}_e . \hat{\mathbf{n}}$ with $\nabla n_s . \hat{\mathbf{n}} = 0$. Here, $\nabla T_e . \hat{\mathbf{n}} = -G_r (T_b - T_a) / K$ is the Robin boundary condition with $T_a$ the ambient room temperature, $T_b$ the extrapolated temperature at the boundary, $G_r = 10^6$ W/m$^2$K the surface heat transfer rate, and $K = 40$ W/mK the thermal conductivity.

The spin polarization current density contains the usual drift and diffusion terms, where $P$ is the current spin polarization in a ferromagnet (FM), with $\mathbf{m}$ the magnetic spin direction, as:

$$\mathbf{J}_S = -\frac{\mu_B}{e} P \mathbf{J}_C \otimes \mathbf{m} - D_e \nabla \mathbf{S} \tag{4}$$

$\mathbf{S}$ is the spin accumulation which obeys a continuity relation using the divergence of the spin current. Evaluating $\nabla \cdot \mathbf{J}_S$ involves obtaining a gradient in $\mathbf{m}$, which is normally used with micromagnetic models [36,37]. With ASD special consideration must be given since the angles between neighbouring atomistic moments can be very large, particularly at higher temperatures. To resolve this, the spin accumulation can be solved with a finer discretization than the atomistic lattice constant. The results converge when the $\mathbf{S}$ cellsize is refined to be at least 3 times smaller than the atomistic lattice constant. In particular, in the slow current transient regime where the spin accumulation response time is relatively negligible (laser pulses longer than a few ps), the spin torque due to spin-polarized current transients between magnetic layers must tend to the Slonczewski spin torque [38], and this is reproduced by the self-consistent spin torque from the ASDD model as shown in the Supplementary Information. The continuity equation for **S** is:

$$\frac{\partial \mathbf{S}}{\partial t} = -\nabla \cdot \mathbf{J}_S - \chi_{sp}\frac{\partial \mathbf{m}}{\partial t} - D_e\left[\frac{\mathbf{S}}{\lambda_{sf}^2} + \frac{\mathbf{S}\times\mathbf{m}}{\lambda_J^2} + \frac{\mathbf{m}\times(\mathbf{S}\times\mathbf{m})}{\lambda_\varphi^2}\right] \tag{5}$$

Here, the second term is due to spin pumping, introduced as discussed in Ref. [39], where $\chi_{sp} = \mu_B \hbar \sigma / e^2 \lambda_J^2$. The last term includes: i) longitudinal relaxation with spin-flip length $\lambda_{sf}$, ii) precessional-like relaxation with exchange rotation length $\lambda_J$, and iii) damping-like relaxation with spin dephasing length $\lambda_\varphi$. Here we use $P = 0.6$, $\lambda_J = 4$ nm, $\lambda_\varphi = 2$nm, $\lambda_{sf} = 10$ nm, and $D_e = 10^{-4}$ m$^2$/s. Through the transverse relaxation terms the spin accumulation exerts a reciprocal spin torque on the magnetic spins – the self-consistent spin torque – given as:

$$\mathbf{T}_S = -\frac{D_e}{\lambda_J^2}\mathbf{m}\times\mathbf{S} - \frac{D_e}{\lambda_\varphi^2}\mathbf{m}\times(\mathbf{m}\times\mathbf{S}) \tag{6}$$

While in the slow current transient regime the self-consistent spin torque in a spin valve geometry tends to the Slonczewski spin torque, $\mathbf{T}_S = \eta(\mu_B / e)(J_C / d_{FL})\mathbf{m}\times(\mathbf{m}\times\mathbf{p})$, the latter does not accurately model AOS in the ultrafast regime, even when the current density is computed using Equation (1). The spin accumulation response time becomes important in the ultrafast regime, where AOS depends on the timing between the spin torque and demagnetization and remagnetization processes. Moreover, the macrospin approximation used to obtain the Slonczewski spin torque, where the free layer (FL) has spin direction **m** and thickness $d_{FL}$, and the reference layer (RL) has spin direction **p**, does not hold at higher laser fluences where multi-domain structures can be created. Finally, the STT polarization, $\eta$, is dependent on the layer thicknesses and spin transport parameters [40], and in the ultrafast regime also depends on the pulse width as shown in the Supplementary Information. Such non-local effects are naturally included when the self-consistent spin torque is used in conjunction with ASD. Equation (6) is included in the stochastic atomistic Landau Lifshitz Gilbert (sLLG) equation, where $\alpha$ is the damping constant, $a$ is the atomistic lattice constant and $\mu_S$ the corresponding magnetic moment:

$$\frac{\partial \mathbf{m}}{\partial t} = -\gamma \mathbf{m} \times (\mathbf{H} + \mathbf{H}_{th}) + \alpha \mathbf{m} \times \frac{\partial \mathbf{m}}{\partial t} + \frac{a^3}{\mu_S} \mathbf{T}_S \tag{7}$$

The stochastic thermal field $\mathbf{H}_{th}$ follows a Gaussian distribution with standard deviation $H_{th}^{(\sigma)} = \sqrt{2\alpha k_B T_e / \gamma \mu_0 \mu_S \Delta t}$, where $\Delta t$ is the integration time-step. The effective field $\mathbf{H}$ includes contributions from: i) exchange interaction $\mathbf{H}_{ex} = (J / \mu_0 \mu_S) \sum_{j \in N} \mathbf{m}_j$, where $J$ is the exchange energy with sum running over nearest neighbours, ii) uniaxial anisotropy $\mathbf{H}_{an} = (2K_u / \mu_0 \mu_S)(\mathbf{m}.\mathbf{e}_A)\mathbf{e}_A$, where $K_u$ is the anisotropy energy with symmetry axis direction $\mathbf{e}_A$, and iii) dipole-dipole interaction $\mathbf{H}_{d-d,i} = \mu_S \sum_{i \neq j} [3(\mathbf{m}_j.\hat{\mathbf{r}}_{ij})\hat{\mathbf{r}}_{ij} - \mathbf{m}_j] / 4\pi r_{ij}^3$ at spin $i$, where the sum runs over all other spins such that $\hat{\mathbf{r}}_{ij} r_{ij}$ is the distance vector from spin $i$ to spin $j$. Here we use $\mu_S = 6.4\,\mu_B$ (corresponding to a ground state magnetization of $M_S = 475$ kA/m with $a = 5$ Å), $J = 6.8 \times 10^{-21}$ J, $K_u = 1.25 \times 10^{-22}$ J, and $\alpha = 0.1$. With the set exchange interaction energy the Curie temperature is ~710 K, as determined using Monte Carlo simulations [41]. The $a = 5$ Å lattice constant for a simple cubic atomistic lattice was chosen for computational efficiency, with the resultant magnetic moment $\mu_S = M_S / a^3$, since the model is applied here to understand the ultrafast AOS mechanisms in a generic spin valve structure.

Finally, the temperature is computed using a two-temperature model, with $T_l$ the lattice temperature, $C_e = 800$ J/kgK and $C_l = 40$ J/kgK the electron and lattice specific heat capacities respectively, $\rho = 8740$ kg/m$^3$ the mass density, and $G_e = 2 \times 10^{18}$ W/m$^3$K the electron-lattice heat transfer rate:

$$\begin{aligned} C_e \rho \frac{\partial T_e}{\partial t} &= K \nabla^2 T_e - G_e (T_e - T_l) + Q \\ C_l \rho \frac{\partial T_l}{\partial t} &= G_e (T_e - T_l) \end{aligned} \tag{8}$$

Another contribution which could be considered here is the Onsager reciprocal term to the Seebeck effect, namely a Peltier term, $\nabla\Pi.\mathbf{J}_C$, with Peltier coefficient $\Pi = TS_c$. This acts as a secondary source term in the heat equation. However, we estimate its contribution to be at least 2 orders of magnitude smaller compared to the source term due to the laser pulse, $Q$, and is ignored here for simplicity.

In Equation (1) we neglected the long-ranged electrostatic interaction arising due to the excess charge distribution. This can be calculated from Maxwell's first equation, $\nabla.\mathbf{E} = n_e / \varepsilon_0$, resulting in the additional free current density contribution $\mathbf{J}_f = \sigma\mathbf{E}$. This is a secondary contribution, which we estimate to be 2 – 3 orders of magnitude smaller than that due to constitutive relations included in Equation (1) – numerical details are given in the Supplementary Information. A combined Maxwell's equations and spin dynamics solver, whilst computationally challenging, would be an important extension of the current model. This is particularly useful for helicity-dependent AOS, where optical magnetic fields are generated due to the circular polarization. This is beyond the scope of the current work where we restrict the laser pulse to linear polarization.

To summarize, the ASDD model comprises the dynamics Equations (2), (3), (5), (7) and (8). The sLLG equation is solved using the RK4 method with $\Delta t$ = 5 fs, and the remaining equations are solved using the forward-time centered-space scheme with a time-step of 0.2 fs. These are discretized using the lattice constant $a$, except for $\mathbf{S}$ which uses a 4-fold refinement per lattice cell. To allow large-scale simulations the ASDD model was implemented in BORIS [42] using multi-GPU acceleration [43].

## Results

The structure investigated here is a generic tri-layer spin valve, FL(2 nm)/SL($d_{SL}$)/RL($d_{RL}$), with the material parameters given above, as sketched in Figure 1(b) inset. The spacer layer (SL) thickness is varied between $d_{SL} = 0.4\lambda_{sf}$ and $d_{SL} = 2.4\lambda_{sf}$. The RL thickness is fixed as $d_{RL}$ = 15 nm or 10 nm, and the FL thickness is fixed as $d_{FL}$ = 2 nm throughout.

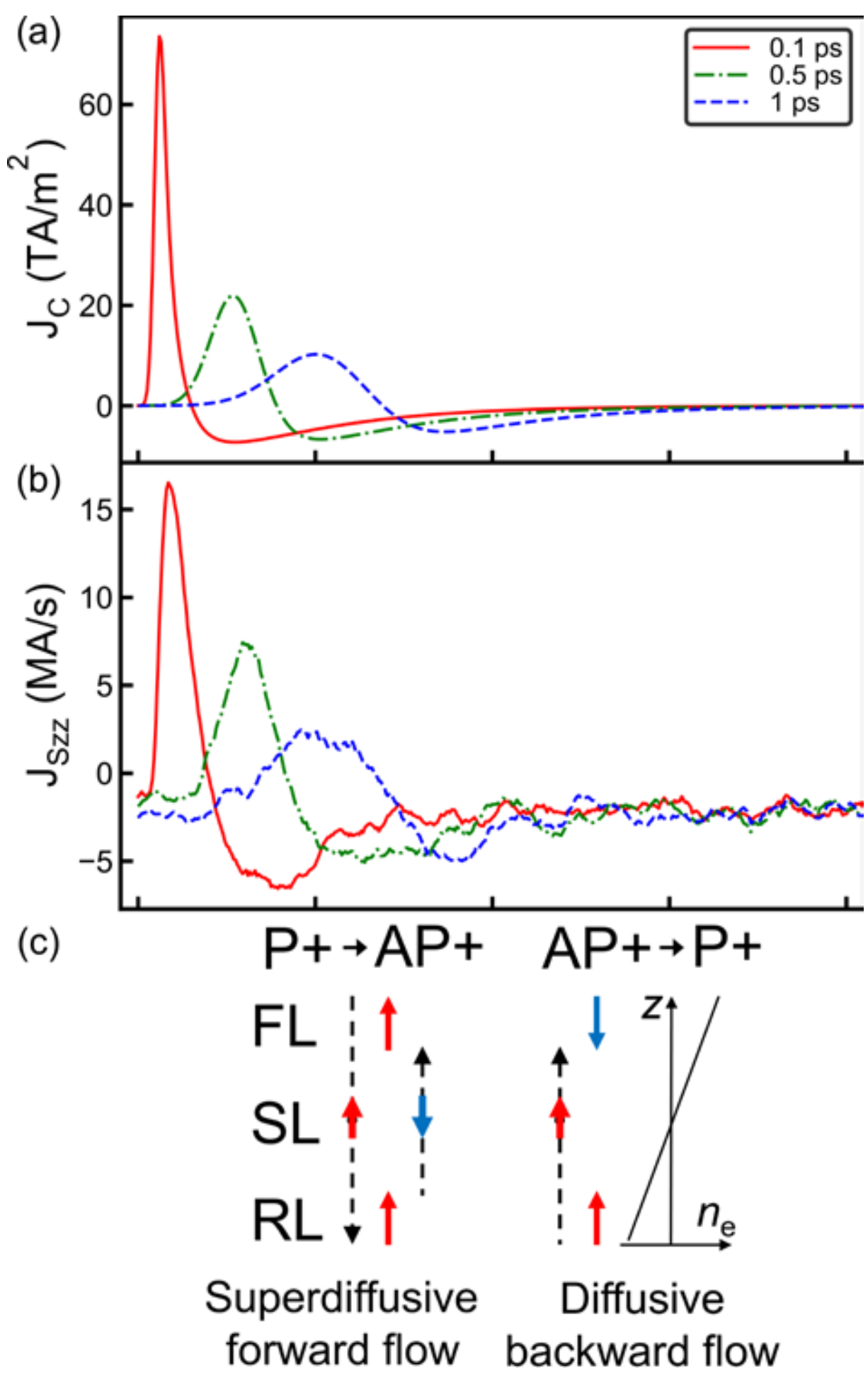

Figure 1 – Laser-induced current transients, using $d_{RL}$ = 15 nm and $d_{SL} = 0.8\lambda_{sf}$. (a) Spin-polarized current transients in the FL for 3 mJ/cm$^2$ fluence and low superdiffusive regime. The positive current transient is due to superdiffusive forward flow of non-equilibrium electrons, and the negative current transient is due to diffusive backward flow of excess electrons as the charge distribution re-equilibrates. (b) Pure spin current transients in the SL, pumped by the RL, for 5 mJ/cm$^2$ fluence and high spin pumping regime, showing the *z* component of the *z* direction spin currents. The positive and negative current transients are due to demagnetization

and remagnetization of the RL, respectively. (c) Illustration of switching mechanisms in the spin valve, with P+ to AP+ switching occurring on the superdiffusive forward flow due to accumulation of minority spins at the RL, and AP+ to P+ switching occurring on the diffusive backward flow, repolarized by the RL, as excess charges re-equilibrate.

For a broader investigation into the switching characteristics we define low and high superdiffusive regimes with $v_e = 0.25\times10^6$ m/s and $v_e = 1.00\times10^6$ m/s, respectively. Similarly, we define low and high spin pumping regimes with $\chi_{sp}$ = 2.4 kA/m ($\sigma = 10^6$ S/m) and $\chi_{sp}$ = 24 kA/m ($\sigma = 10^7$ S/m), respectively. The laser pulse is applied to the spin valve from the FL surface, with $Q(z,t) = Q_0 \exp(-(d_{RL} + d_{SL} + d_{FL} - z)/d - 8(t - \tau_{pw})^2 / \tau_{pw}^2)$, where $\tau_{pw}$ is the pulse width, $Q_0$ is the maximum power density, and $d$ = 4 nm is the optical absorption length. The laser fluence is obtained from the power density by integrating over $z$ and $t$.

Example current transients are shown in Figure 1. In particular, Figure 1(a) shows the charge current transients. The initial positive current transient is due to superdiffusive forward flow of electrons (from top to bottom). These non-equilibrium electrons are excited by the laser pulse, superdiffuse away from temperature gradients, and have a short relaxation time on the ultrafast time-scale. Thus, as the pulse width is increased, the amplitude of the positive current transient decreases rapidly. The amplitude of the spin-polarized currents in Figure 1(a) are in order-of-magnitude agreement with experimental observations of hot electron transport [35]. A consequence of this forward flow of electrons is creation of an excess charge density. As electrons superdiffuse away from the FL, a net positive charge density is created at the top, and as electrons accumulate at the bottom of the RL, a net negative charge density is created there. Example charge density and temperature profiles are shown in the Supplementary Information. Thus, after the laser pulse, a delayed diffusive backward flow of electrons is generated as the charge density re-equilibrates, resulting in the negative current transients in Figure 1(a). The net charge displaced, obtained by integrating the current transients in Figure 1(a), is zero, as expected from charge conservation. Since the diffusive backward flow of electrons is a slower process, the negative current transient amplitude is smaller, and as the pulse width increases the forward and backward transient amplitudes tend to the same value. Thus, for shorter pulses and not too high fluences, the spin torque due to the superdiffusive forward transient dominates, and for longer pulses or higher fluences the diffusive backward transient becomes increasingly important. As shown in the Supplementary Information, the self-consistent spin torque from Equation (6) is an ultrafast Slonczewski spin torque. The main difference between them is a time lag of the self-consistent spin torque relative to the Slonczewski spin torque, due to the finite spin accumulation response time, the latter being approximated as directly proportional to the spin-polarized charge current. While for longer pulse widths (a few ps) this time lag is relatively negligible, being of the order of a few hundred fs, in the ultrafast regime it is not. Nevertheless, the switching mechanism in these metallic spin valves is broadly the same as that described by the well-known Slonczewski spin torque. That is, P+ to AP+ switching of the FL occurs on the positive current transient due to accumulation of minority spins at the RL, as spin-polarized electrons flow from the FL towards the RL. AP+ to P+ switching occurs on the negative current transient, as electrons flow from the RL towards the FL, transferring the spin polarization of the RL. Since the thicker RL has a much lower average temperature only the FL is switched at lower fluences. At higher fluences the RL can also be switched. In agreement with Ref. [23], a requirement for switching (of either the FL or RL) is complete demagnetization of the switched layer. Example temperature transients, magnetization switching dynamics, and spin torque temperature dependences are shown in the Supplementary Information. It is known that superdiffusive currents contribute to ultrafast demagnetization [27]. Here ultrafast demagnetization is modelled via the phenomenological 2TM of Equation (8), coupled to the ASD model, with the effect of superdiffusive currents captured via the STT of Equation (6). An additional effect of this s-d exchange coupling is a decrease of the magnetic moments, as shown in Ref. [27], which also contributes to ultrafast demagnetization. Whilst this is important for exact numerical reproduction of experimental works, in particular demagnetization and remagnetization rates, the inclusion of this effect within the ASD model is beyond the scope of the current work.

The Seebeck effect with a positive coefficient also contributes to the charge current transients. However, since the Seebeck coefficient in metals is small, even with an upper estimate of $S_c$ = 10 μV/K, and $\sigma = 10^7$ S/m, the charge current transient amplitude is an order of magnitude smaller for the Seebeck effect compared to the superdiffusive mechanism, and may be neglected as shown in the Supplementary Information. This conclusion is in agreement with the experimental observations in Ref. [25].

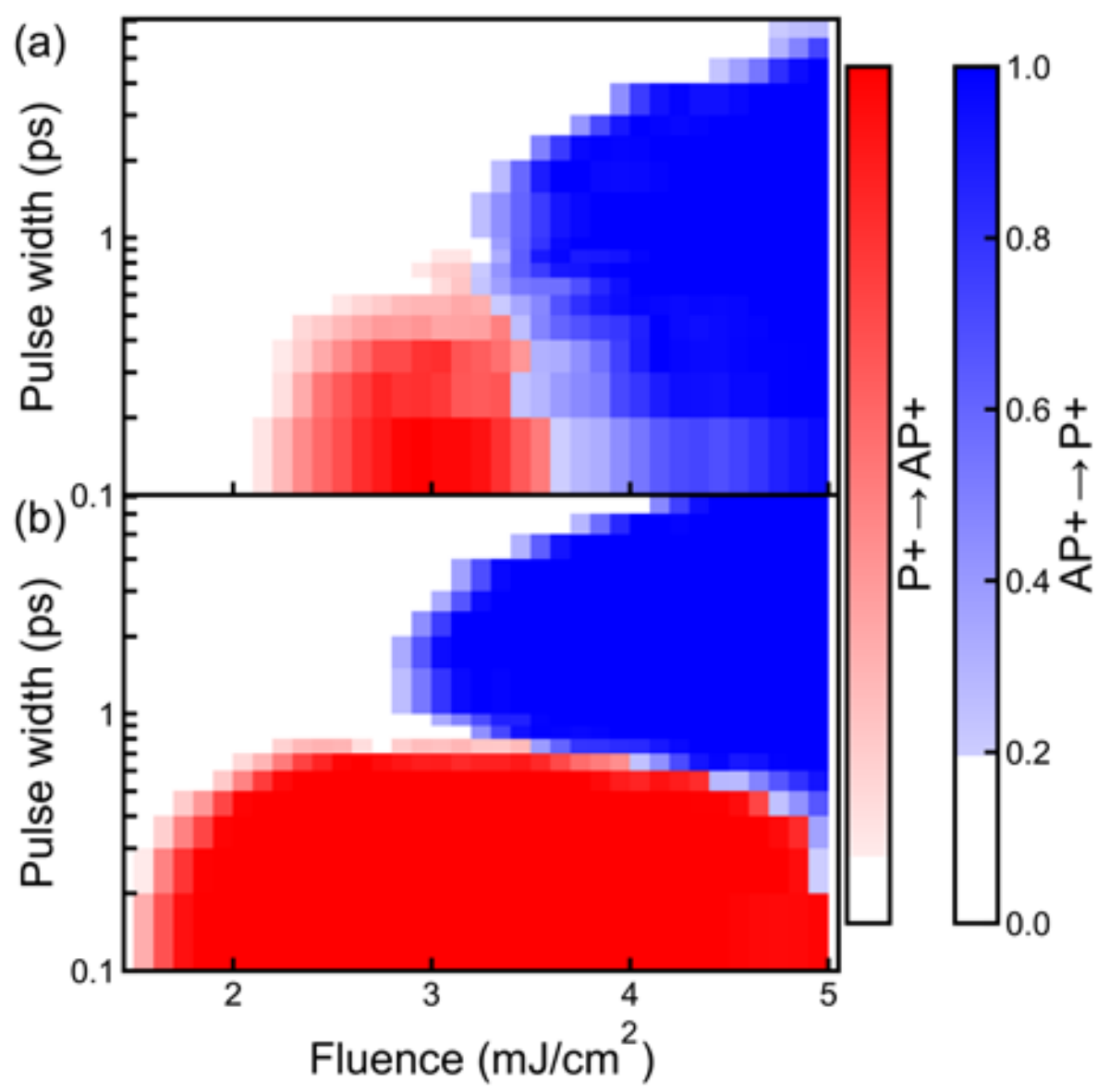

Figure 2 – Switching probability as a function of a pulse width and laser fluence, using $d_{RL}$ = 15 nm and $d_{SL}$ = $0.8\lambda_{sf}$. The red and blue scales indicate P+ to AP+, respectively AP+ to P+, switching probability. (a) Low superdiffusive regime with $v_e$ = $0.25\times10^6$ m/s, and (b) high superdiffusive regime with $v_e$ = $1.00\times10^6$ m/s.

Another source of current transients is demagnetization and remagnetization, which generate pure spin currents through spin pumping. Examples are shown in Figure 1(b), plotting the spin current obtained in the SL due to the RL. With the RL magnetization direction indicated in Figure 1(b), during its demagnetization a positive spin current is pumped in the SL. This favors the P+ orientation. During remagnetization of the RL a negative spin current is pumped in the SL. This favors the AP+ orientation and, as shown in the Supplementary Information, sufficiently strong spin pumping alone can result in P+ to AP+ switching. However, AP+ to P+ switching cannot obtained with spin pumping alone, in agreement with Ref. [21].

The switching probabilities, as a function of laser fluence and pulse width, are shown in Figure 2. These are calculated starting from the P+ and AP+ states using 25 nm side squares, repeated over 9 events, for the low and high superdiffusive regimes, in combination with the low spin pumping regime. High spin pumping regime was also simulated. For short pulses and low fluences P+ to AP+ switching is obtained, as the spin torque due to the forward flow transient dominates. Here, the spin torque due to the backward flow transient is not sufficiently strong to obtain a P+ orientation. However, as the fluence is increased, eventually the latter transient becomes more important, resulting in AP+ to P+ switching, also suppressing P+ to AP+ switching. Moreover, as the pulse width is increased, the region in which P+ to AP+ switching is obtained shrinks, and for longer pulses or higher fluences only AP+ to P+ switching is possible. This is due to the decrease of the forward flow transient amplitude with pulse width, as discussed above in reference to Figure 1(a). Increasing the fluence does not help, since the amplitude of the backward flow transient also increases, resulting in predominantly AP+ to P+ switching. We note from very recent experimental results [23] that the P+ to AP+ switching region was extended to longer pulses by the introduction of a Cu capping layer. One effect of a thicker spin valve structure is a change in the balance of the forward and backward flow transient amplitudes, with the latter becoming shallower but longer due to increased diffusion time. Whilst reproduction of these results is beyond the scope of the current work (however, see Figure 5 and associated discussion), we speculate that increasing the overall spin valve thickness using a capping layer allows extending the P+ to AP+ switching region to longer pulses, as the AP+ to P+ switching threshold also increases.

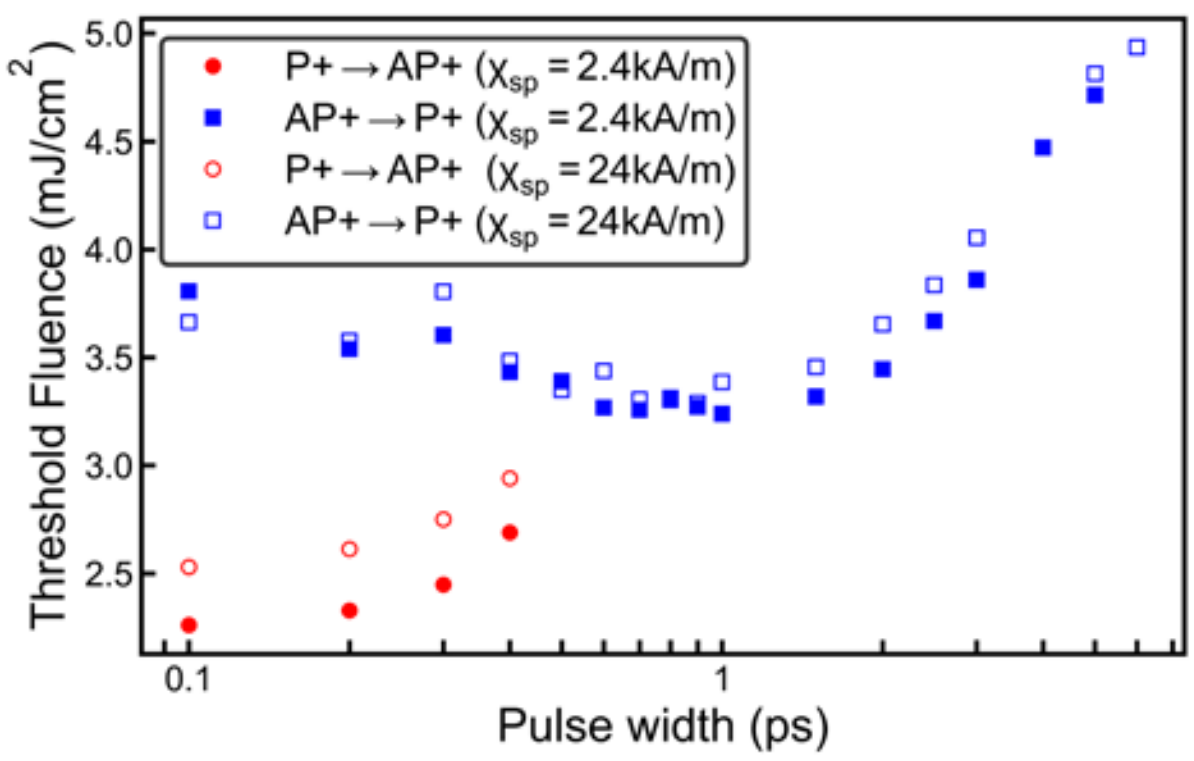

Figure 3 – Threshold fluence as a function of pulse width for P+ to AP+ (circles), and AP+ to P+ switching (squares), respectively. The threshold values are obtained from the low superdiffusive regime results in Figure 2, shown for both low (closed symbols) and high (open symbols) spin pumping regimes.

Threshold fluences for switching are shown in Figure 3, obtained from the results in Figure 2(a) for the low superdiffusive regime, shown for both low and high spin pumping regimes. These results are in agreement with the experimental results in Ref. [21] (Extended Data Fig. 7 in this reference). The exact threshold values are dependent on the material parameters, in particular heat, electron and spin transport, as well as the exact spin valve geometry. For example, the threshold fluences obtained in the strong superdiffusive regime in Figure 2(b) are significantly different than those in the low superdiffusive regime – the P+ to AP+ switching region is expanded due to the stronger superdiffusive forward flow transient, which consequently also impacts the AP+ to P+ switching region. Exact numerical reproduction of experimental results is difficult, and is beyond the scope of this work. These results show that the spin torque due to the spin-polarized current transient is the main mechanism through which magnetization switching is obtained in these ferromagnetic metallic spin valves. The effect of spin pumping is small, even when combined with the low superdiffusive regime. The spin current pumped by the RL during its remagnetization favors an AP+ orientation, and thus could be expected to increase the threshold fluence values for AP+ to P+ switching. However, these are not significantly affected, since the diffusive backward flow spin-polarized current transient is stronger, which largely determines the threshold fluence values. There is a small shift in the threshold fluence values for P+ to AP+ switching with higher spin pumping, as seen in Figure 3. This is due to the spin current pumped by the RL during demagnetization, which favors the P+ orientation.

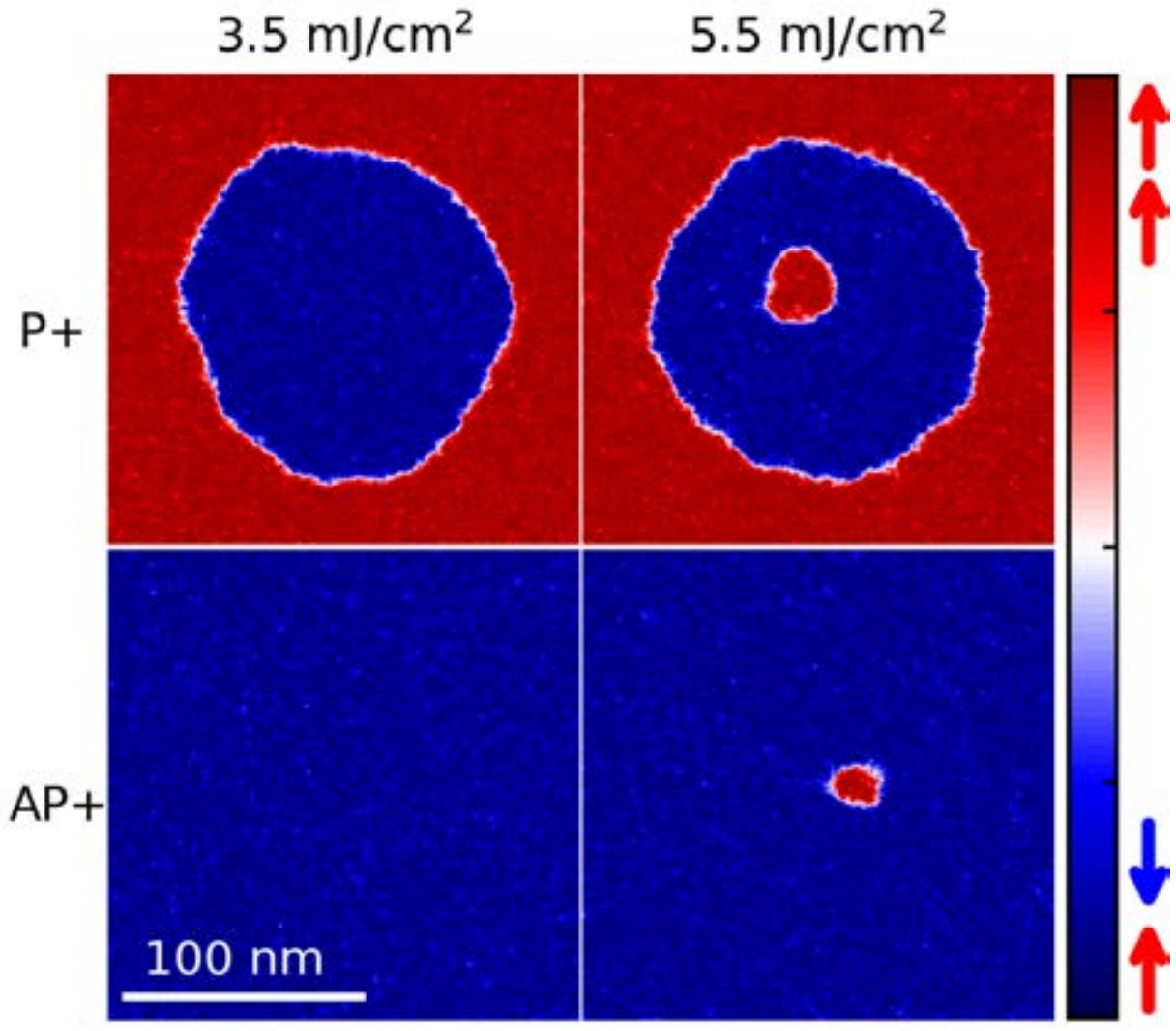

Figure 4 – AOS using a Gaussian laser spot for P+ and AP+ starting states, and different laser fluences, using $d_{RL}$ = 15 nm and $d_{SL} = 0.8\lambda_{sf}$ in the high superdiffusive regime. For low laser fluence, P+ switches to AP+ under the laser spot, however the AP+ starting state is not affected. P+ to AP+ switching is driven by the superdiffusive forward current transient. For high laser fluence, where AP+ to P+ switching is possible due to an increase in the diffusive backward current transient, a small domain in the P+ state is found at the center for both the P+ and AP+

starting states. This is due to a combination of the Gaussian beam shape and in-plane diffusion of excess electrons, resulting in stronger diffusive backward current transient at the center of the laser spot.

The experimental observations in Ref. [21] have revealed further details in the switching process, namely creation of multi-domain structures at higher fluences, as well as creation of central smaller P+ regions in both AP+ to P+ and P+ to AP+ switching. Next, we turn our attention to spatially resolved simulations using a circular laser spot with a Gaussian intensity profile. Selected results for 100 fs pulse width are shown in Figure 4. A full set of spatially resolved simulations, as a function of fluence and pulse width up to 2 ps, are shown in the Supplementary Information. For the lower fluence of 3.5 mJ/cm$^2$, P+ to AP+ switching is obtained under the laser spot, and the AP+ initial state is unaffected. This fluence is above the threshold value for P+ to AP+ switching, but also below that for AP+ to P+ switching. For the higher fluence of 5.5 mJ/cm$^2$ P+ to AP+ switching still occurs, however a smaller central P+ region is also obtained, as seen in Figure 4. Initially, the entire region under the laser spot is switched to the AP+ configuration, under the effect of the forward flow transient. The backward flow transient now has sufficient amplitude to switch the configuration back to P+, as this fluence exceeds the threshold value. However, the transient amplitudes have a spatial dependence, being stronger at the center. This is partly due to the Gaussian beam profile. For the diffusive backward transient another source of spatial variation is in-plane diffusion of excess charge, which again results in a stronger amplitude at the center. Due to this, results similar to those in Figure 4 may be obtained even with a top-hat beam, although the Gaussian beam profile is more realistic. Similarly, AP+ to P+ switching is obtained at the higher fluence, but only in a much smaller area where the backward flow transient amplitude exceeds the required threshold. These results are in agreement with those in Ref. [21] (see Fig. 2 in this reference), providing strong evidence for the switching mechanism described here.

Furthermore, as in Ref. [21], multi-domain structures are obtained at higher fluences and longer pulses, which can be in the FL only, but also in both the FL and RL at even higher fluences (see Supplementary Information). For the structures investigated here, P+ to AP+ switching, with or without a central smaller P+ region, is only obtained for pulse widths up to 0.6 ps. For longer pulse widths, the reduced superdiffusive forward flow transient results in incomplete P+ to AP+ switching, appearing as a multi-domain structure. Also, for larger fluences the central P+ region grows to the point where it becomes comparable in size to the initial switched AP+ region, which also forms multi-domain structures. At even higher fluences the RL is also switched, generating multi-domain structures in both FL and RL; this is also true with an AP+ initial state.

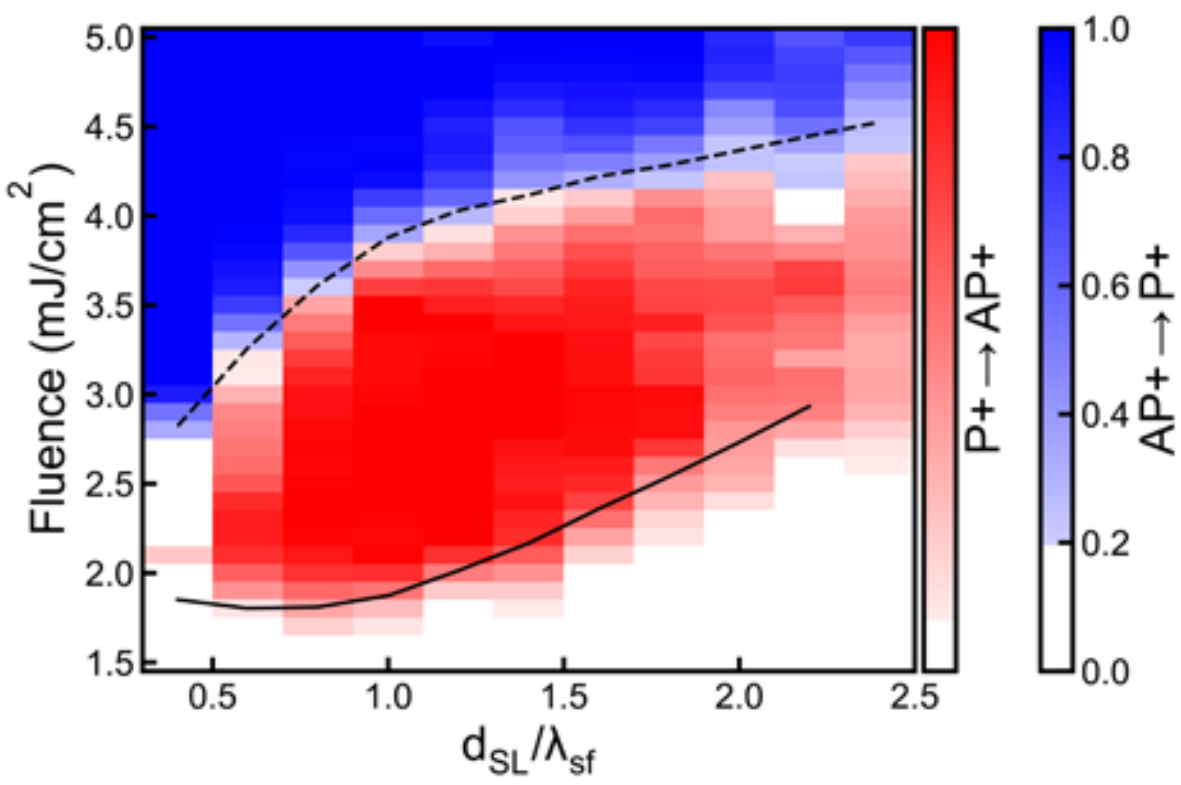

Figure 5 – Switching probability as a function of laser fluence and SL thickness, using $d_{RL}$ = 10 nm and 0.1 ps pulse width. The red and blue scales indicate P+ to AP+, respectively AP+ to P+, switching probability, with the black lines indicated the switching threshold fluences.

Finally, we investigate the effect of varying the SL thickness. A switching probability map is shown in Figure 5 for 0.1 ps pulse width, as a function of laser fluence and SL thickness. If the SL thickness is too small, P+ to AP+ switching is not possible. This was experimentally observed in perpendicular anisotropy spin-valve structures [22], where below a certain Cu spacer layer thickness P+ to AP+ switching was not possible using a single laser pulse, as the ferromagnetic RKKY interlayer exchange coupling plays an important role. While for the thicker spacer layers used here inclusion of RKKY exchange is beyond the scope of the current study, we point out that P+ to AP+ switching is also prohibited for thin spacer layers due to the nature of laser-induced current transients. Here, this is due to the reduced overall thickness of the spin valve, resulting in larger gradients of excess charge density accumulated at the top and bottom surfaces. This compression of excess charge density, and shorter diffusion time, increases the amplitude of the backward flow transient; however, the amplitude of the forward flow transient is largely unaffected. Thus, AP+ to P+ switching is obtained at lower fluences in spin valves with a thinner SL. Increasing the SL thickness primarily results in an increase in the AP+ to P+ switching threshold, as the backward flow transient becomes shallower, but longer. Thus, up to $d_{SL} \cong \lambda_{sf}$ the threshold for P+ to AP+ changes only slightly, whereas that for AP+ to P+ increases rapidly. This was also observed in Ref. [21], and the results in Figure 5 provide an explanation for these experimental findings. The spin torques are also dependent on $d_{SL}$, since the spin accumulation between RL and FL decays exponentially in the SL. Further increasing $d_{SL}$ results in increase in the threshold fluences for both types of switching, and eventually the switching probabilities decrease to the point where deterministic switching is no longer possible.

## Conclusion

AOS in ferromagnetic spin valves was investigated here using an atomistic spin drift-diffusion dynamics model. This combines an atomistic spin dynamics model, with calculation of a self-consistent spin torque based on a spin drift-diffusion model. The latter includes contributions from superdiffusive hot electrons, spin pumping and spin-dependent Seebeck effects. It is found the magnetization switching in such spin valves is governed by the laser-induced current transients and associated spin torques. In particular, a forward flow of electrons is generated away from the laser-exposed surface due to superdiffusive hot electrons. These carry spin angular momentum from the FL, resulting in accumulation of minority spins at the RL. The accumulation of these minority spins drives P to AP switching. As electrons accumulate at the bottom of the spin valve, following the initial superdiffusive forward flow, the charge distribution is found in a non-equilibrium state immediately after the laser pulse, with a net negative charge at the bottom, and a net positive charge at the surface of the spin valve. This results in a diffusive backward flow of electrons as the charge is driven back into equilibrium during the relaxation phase. The backward flowing electrons, repolarized by the RL, exert a spin torque on the FL, and drive AP to P switching. Since the amplitudes of these laser-induced current transients depend on both the laser fluence and pulse width, we are able to reproduce experimentally observed fluence and pulse-width-dependent P to AP and AP to P magnetization switching, as well as formation of multi-domain structures and dependence on spacer layer thickness. While spin pumping and spin-dependent Seebeck effects contribute to the overall laser-induced current transients, we find their role is minor compared to the superdiffusive hot electron generation mechanism, and are not essential to explain the AOS observed in these structures.

## References

[1] I. Radu, K. Vahaplar, C. Stamm, T. Kachel, N. Pontius, H. A. Dürr, T. A. Ostler, J. Barker, R. F. L. Evans, R. W. Chantrell *et al.*, Transient ferromagnetic-like state mediating ultrafast reversal of antiferromagnetically coupled spins, Nature **472**, 205 (2011).

[2] T.A. Ostler, J. Barker, R.F.L. Evans, R.W. Chantrell, U. Atxitia, O. Chubykalo-Fesenko, S. El Moussaoui, L. Le Guyader, E. Mengotti, L.J. Heyderman *et al.*, Ultrafast heating as a sufficient stimulus for magnetization reversal in a ferrimagnet, Nat. Commun. **3**, 666 (2012).

[3] M. Verges, W. Zhang, Q. Remy, Y. Le-Guen, J. Gorchon, G. Malinowski, S. Mangin, M. Hehn, and Julius Hohlfeld, Extending the scope and understanding of all-optical magnetization switching in Gd-based alloys by controlling the underlying temperature transients, Phys. Rev. Appl. **21**, 044003 (2024).

[4] Y. Xu, M. Deb, G. Malinowski, M. Hehn, W. Zhao and S. Mangin, Ultrafast Magnetization Manipulation Using Single Femtosecond Light and Hot-Electron Pulses, Adv. Mater. **29**, 1703474 (2017).

[5] N. Bergeard, V. López-Flores, V. Halté, M. Hehn, C. Stamm, N. Pontius, E. Beaurepaire, and C. Boeglin, Ultrafast angular momentum transfer in multisublattice ferrimagnets, Nat. Commun. **5**, 3466 (2014).

[6] M. L. M. Lalieu, M. J. G. Peeters, S. R. R. Haenen, R. Lavrijsen, and B. Koopmans, Deterministic all-optical switching of synthetic ferrimagnets using single femtosecond laser pulses, Phys. Rev. B **96**, 220411 (2017).

[7] P. Li, T. J. Kools, H. Pezeshki, J. M. B. E. Joosten, J. Li, J. Igarashi, J. Hohlfeld, R. Lavrijsen, S. Mangin, G. Malinowski *et al.*, Picosecond all-optical switching of Co/Gd–based synthetic ferrimagnets, Phys. Rev. B **111**, 064421 (2025).

[8] C. R. J. Sait, M. Dąbrowski, J. N. Scott, W. R. Hendren, D. G. Newman, A. T. N'Diaye, C. Klewe, P. Shafer, G. van der Laan, P. S. Keatley *et al.*, Unidirectional multipulse helicity-independent all-optical switching in [Ni/Pt] based synthetic ferrimagnets, Phys. Rev. B **109**, 134417 (2024).

[9] J. Gorchon, C. -H. Lambert, Y. Yang, A. Pattabi, R. B. Wilson, S. Salahuddin, and J. Bokor, Single shot ultrafast all optical magnetization switching of ferromagnetic Co/Pt multilayers, Appl. Phys. Lett. **111**, 042401 (2017).

[10] S. Iihama, Y. Xu, M. Deb, G. Malinowski, M. Hehn, J. Gorchon, E. E. Fullerton, and S. Mangin, Single-Shot Multi-Level All-Optical Magnetization Switching Mediated by Spin Transport, Adv. Mater. **30**, 1804004 (2018).
[11] Q. Remy, J. Igarashi, S. Iihama, G. Malinowski, M. Hehn, J. Gorchon, J. Hohlfeld, S. Fukami, H. Ohno, and S. Mangin, Energy Efficient Control of Ultrafast Spin Current to Induce Single Femtosecond Pulse Switching of a Ferromagnet, Adv. Sci. **7**, 2001996 (2020).
[12] Q. Remy, J. Hohlfeld, M. Vergès, Y. Le Guen, J. Gorchon, G. Malinowski, S. Mangin, and M. Hehn, Accelerating ultrafast magnetization reversal by non-local spin transfer, Nat. Commun. **14**, 445 (2023).
[13] D. Gupta, M. Pankratova, M. Riepp, M. Pereiro, B. Sanyal, S. Ershadrad, M. Hehn, N. Pontius, C. Schüßler-Langeheine, R. Abrudan *et al.*, Tuning ultrafast demagnetization with ultrashort spin polarized currents in multi-sublattice ferrimagnets, Nat. Commun. **16**, 3097 (2025).
[14] J. -X. Lin, Y. Le Guen, J. Hohlfeld, J. Igarashi, Q. Remy, J. Gorchon, G. Malinowski, S. Mangin, T. Hauet, and M. Hehn, Femtosecond-laser-induced reversal in in-plane-magnetized spin valves, Phys. Rev. Appl. **22**, 044051 (2024).
[15] Y. Peng, D. Salomoni, G. Malinowski, W. Zhang, J. Hohlfeld, L. D. Buda-Prejbeanu, J. Gorchon, M. Vergès, J. X. Lin, D. Lacour *et al.*, In-plane reorientation induced single laser pulse magnetization reversal, Nat. Commun. **14**, 5000 (2023).
[16] D. Salomoni, Y. Peng, L. Farcis, S. Auffret, M. Hehn, G. Malinowski, S. Mangin, B. Dieny, L. D. Buda-Prejbeanu, R. C. Sousa *et al.*, Field-Free All-Optical Switching and Electrical Readout of Tb/Co-Based Magnetic Tunnel Junctions, Phys. Rev. Appl. **20**, 034070 (2023).
[17] A. J. Schellekens, N. de Vries, J. Lucassen, and B. Koopmans, Exploring laser-induced interlayer spin transfer by an all-optical method, Phys. Rev. B **90**, 104429 (2014).
[18] D. Rudolf, C. La-O-Vorakiat, M. Battiato, R. Adam, J. M. Shaw, E. Turgut, P. Maldonado, S. Mathias, P. Grychtol, H. T. Nembach *et al.*, Ultrafast magnetization enhancement in metallic multilayers driven by superdiffusive spin current, Nat. Commun. **3**, 1037 (2012).
[19] G. Malinowski, F. Dalla Longa, J. Rietjens, P. V. Paluskar, R. Huijink, H. J. M. Swagten and B. Koopmans, Control of speed and efficiency of ultrafast demagnetization by direct transfer of spin angular momentum, Nat. Phys. **4**, 855 (2008).

[20] N. Bergeard, M. Hehn, S. Mangin, G. Lengaigne, F. Montaigne, M. L. M. Lalieu, B. Koopmans, and G. Malinowski, Hot-Electron-Induced Ultrafast Demagnetization in Co/Pt Multilayers, Phys. Rev. Lett. **117**, 147203 (2016).
[21] J. Igarashi, W. Zhang, Q. Remy, E. Díaz, J. -X. Lin, J. Hohlfeld, M. Hehn, S. Mangin, J. Gorchon, and G. Malinowski, Optically induced ultrafast magnetization switching in ferromagnetic spin valves, Nat. Mater. **22**, 725 (2023).
[22] J. Igarashi, Y. Le Guen, J. Hohlfeld, S. Mangin, J. Gorchon, M. Hehn, and G. Malinowski, Influence of interlayer exchange coupling on ultrafast laser-induced magnetization reversal in ferromagnetic spin valves.
[23] K. Ishibashi, J. Igarashi, A. Anadón, M. Hehn, Y. Le Guen, S. Iihama, J. Hohlfeld, J. Gorchon, S. Mangin, G. Malinowski, Single-Shot Magnetization Reversal in Ferromagnetic Spin Valves via Heat Control, Phys. Rev. Lett. **135**, 116702 (2025).
[24] G. -M. Choi, B. -C. Min, K. -J. Lee, and D. G. Cahill, Spin current generated by thermally driven ultrafast demagnetization, Nat. Commun. **5**, 4334 (2014).
[25] A. J. Schellekens, K. C. Kuiper, R. R. J. C. de Wit, and B. Koopmans, Ultrafast spin-transfer torque driven by femtosecond pulsed-laser excitation, Nat. Commun. **5**, 4333 (2014).
[26] G. -M. Choi, C. -H. Moon, B. -C. Min, K. -J. Lee, and D. G. Cahill, Thermal spin-transfer torque driven by the spin-dependent Seebeck effect in metallic spin-valves, Nat. Phys. **11**, 576 (2015).
[27] M. Battiato, K. Carva, and P. M. Oppeneer, Superdiffusive spin transport as a mechanism of ultrafast demagnetization, Phys. Rev. Lett. **105**, 027203 (2010).
[28] M. Battiato, K. Carva, and P. M. Oppeneer, Theory of laser-induced ultrafast superdiffusive spin transport in layered heterostructures, Phys. Rev. B **86**, 024404 (2012).
[29] M. Beens, K. A. de Mare, R. A. Duine, and B. Koopmans, Spin-polarized hot electron transport versus spin pumping mediated by local heating, J. Phys. Condens. Matter **35**, 035803 (2023).
[30] M. Beens, R. A. Duine, and B. Koopmans, Modeling ultrafast demagnetization and spin transport: The interplay of spin-polarized electrons and thermal magnons, Phys. Rev. B **105**, 144420 (2022).
[31] M. Beens, R. A. Duine, and B. Koopmans, S-d model for local and nonlocal spin dynamics in laser 1excited magnetic heterostructures, Phys. Rev. B **102**, 054442 (2020).
[32] Q. Remy, Ultrafast magnetization reversal in ferromagnetic spin valves: An s-d model perspective, Phys. Rev. B **107**, 174431 (2023).

[33] E. I. Harris-Lee, J. K. Dewhurst, S. Shallcross, and S. Sharma, Spin vacuum switching, Sci. Adv. **10**, eado6390 (2024).
[34] E. Najafi, V. Ivanov, A. Zewail, and M. Bernardi, Super-diffusion of excited carriers in semiconductors, Nat. Commun. **8**, 15177 (2017).
[35] A. Melnikov, I. Razdolski, T. O. Wehling, E. Th. Papaioannou, V. Roddatis, P. Fumagalli, O. Aktsipetrov, A. I. Lichtenstein, and U. Bovensiepen, Ultrafast Transport of Laser-Excited Spin-Polarized Carriers in Au/Fe/MgO(001), Phys. Rev. Lett. **107**, 076601 (2011).
[36] C. Abert, L. Exl, F. Bruckner, A. Drews, and D. Suess, A three-dimensional spin-difusion model for micromagnetics, Sci. Rep. **5**, 14855 (2015).
[37] S. Lepadatu, Unifed treatment of spin torques using a coupled magnetisation dynamics and three-dimensional spin current solver, Sci. Rep. **7**, 12937 (2017).
[38] J. C. Slonczewski, Current-driven excitation of magnetic multilayers, J. Magn. Magn. Mater. **159**, L1 (1996).
[39] M. Gattringer, C. Abert, F. Bruckner, A. Chumak, and D. Suess, Micromagnetically integrated numerical model of spin pumping based on spin diffusion, Phys. Rev. B **106**, 024417 (2022).
[40] S. Lepadatu and A. Dobrynin, Self-consistent computation of spin torques and magneto-resistance in tunnel junctions and magnetic read-heads with metallic pinhole defects, J. Phys. Condens. Matter **35**, 115801 (2023).
[41] S. Lepadatu, G. McKenzie, T. Mercer, C. R. MacKinnon, and P. R. Bissell, Computation of magnetization, exchange stiffness, anisotropy, and susceptibilities in large-scale systems using GPU-accelerated atomistic parallel Monte Carlo algorithms, J. Magn. Magn. Mater. **540**, 168460 (2021).
[42] S. Lepadatu, Boris computational spintronics — High performance multi-mesh magnetic and spin transport modeling software, J. Appl. Phys. **128**, 243902 (2020).
[43] S. Lepadatu, Accelerating micromagnetic and atomistic simulations using multiple GPUs, J. Appl. Phys. **134**, 163903 (2023).

# Laser-Induced Current Transients in Ultrafast All-Optical Switching of Metallic Spin Valves

## Supplementary Information

Serban Lepadatu[1,*], Mohammed Gija[1], Alexey Dobrynin[2], Kevin McNeill[2], Mark Gubbins[2], Tim Mercer[1], Steven M. McCann[1], Philip Bissell[1]

[1]*Jeremiah Horrocks Institute for Mathematics, Physics and Astronomy, University of Lancashire, Preston PR1 2HE, U.K.*

[2]*Seagate Technology, 1 Disc Drive, Derry, BT48 0BF, U.K.*

[*] SLepadatu@lancashire.ac.uk

## Charge and temperature transients and profiles

Example electron and lattice temperature transients in the FL, in response to pulse widths of 0.1 ps, 0.5 ps, and 1.0 ps, respectively, for the same laser fluence of 3 mJ/cm$^2$, are shown in Figure S1. Shorter pulses at the same fluence result in larger maximum electron temperatures, as expected. The electron heat is quickly transferred to the lattice, resulting in rapid cooling. The electron and lattice temperatures equilibrate with the ambient room temperature on the nanosecond time-scale.

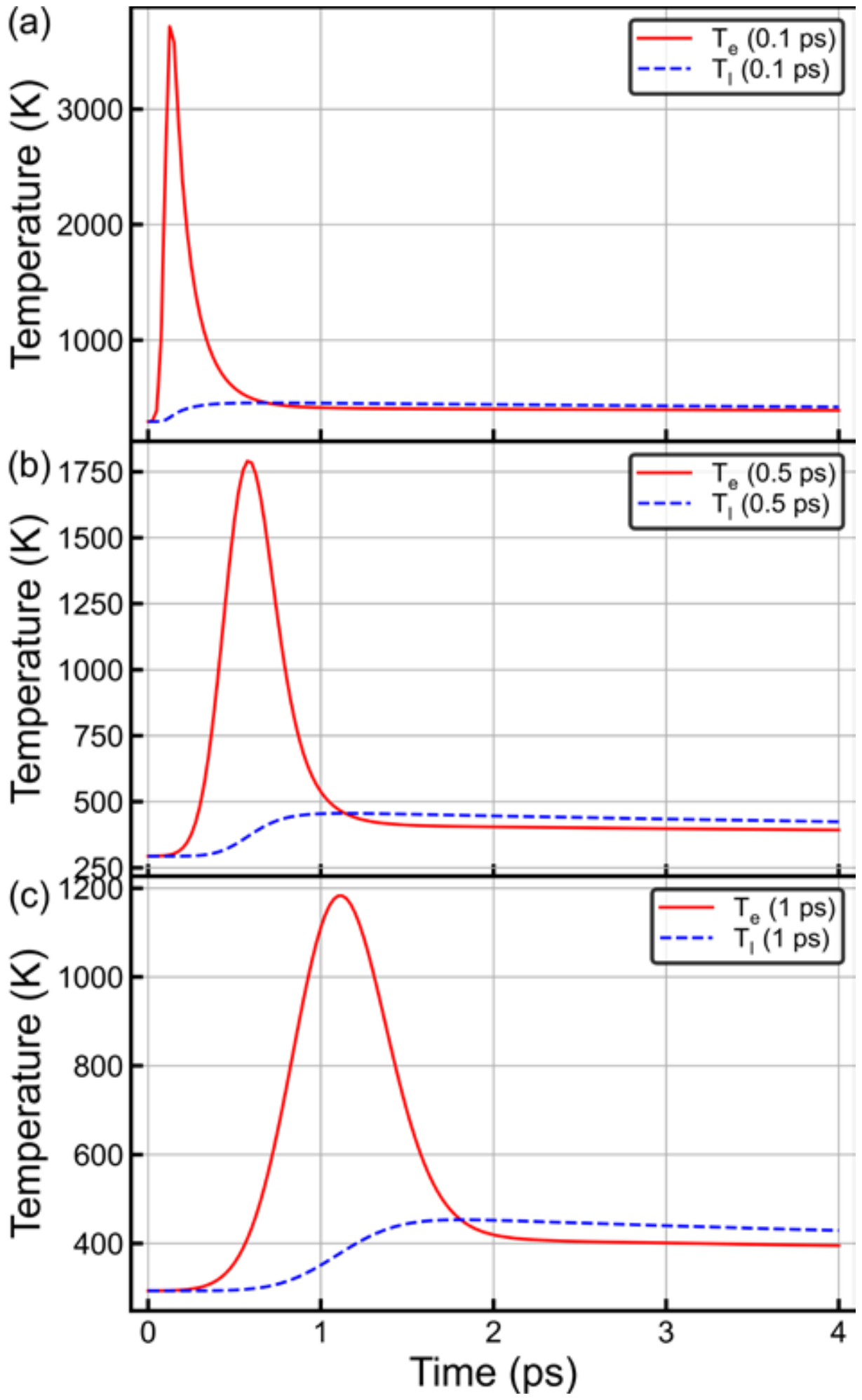

Figure S1 – Two-temperature transients for average electron and lattice temperatures in the FL, corresponding to the results in Figure 1(a) in the main text, with (a) 0.1 ps, (b) 0.5 ps, and (c) 1.0 ps pulse width.

As the laser pulse is applied from the FL side with a Beer-Lambert absorption profile, temperature gradients are generated in the spin valve. Power density absorption profiles, as a function of time for 0.1 ps pulse width at 3 mJ/cm$^2$ fluence, are shown in Figure S2(c). The laser power density is $Q(z,t) = Q_0 \exp(-(d_{RL} + d_{SL} + d_{FL} - z)/d - 8(t - \tau_{pw})^2 / \tau_{pw}^2)$, where $\tau_{pw}$ is the pulse width, $Q_0$ is the maximum power density, and $d$ = 4 nm is the optical absorption length. The laser fluence is obtained from the power density by integrating over $z$ and $t$. Dashed vertical lines in Figure S2 indicate the layer thickness as $d_{RL}$ = 15 nm, $d_{SL}$ = 8 nm, whereas $d_{FL}$ = 2 nm. The resulting electron and lattice temperature profiles, as a function of time, are shown in Figure S2(a), (b), respectively.

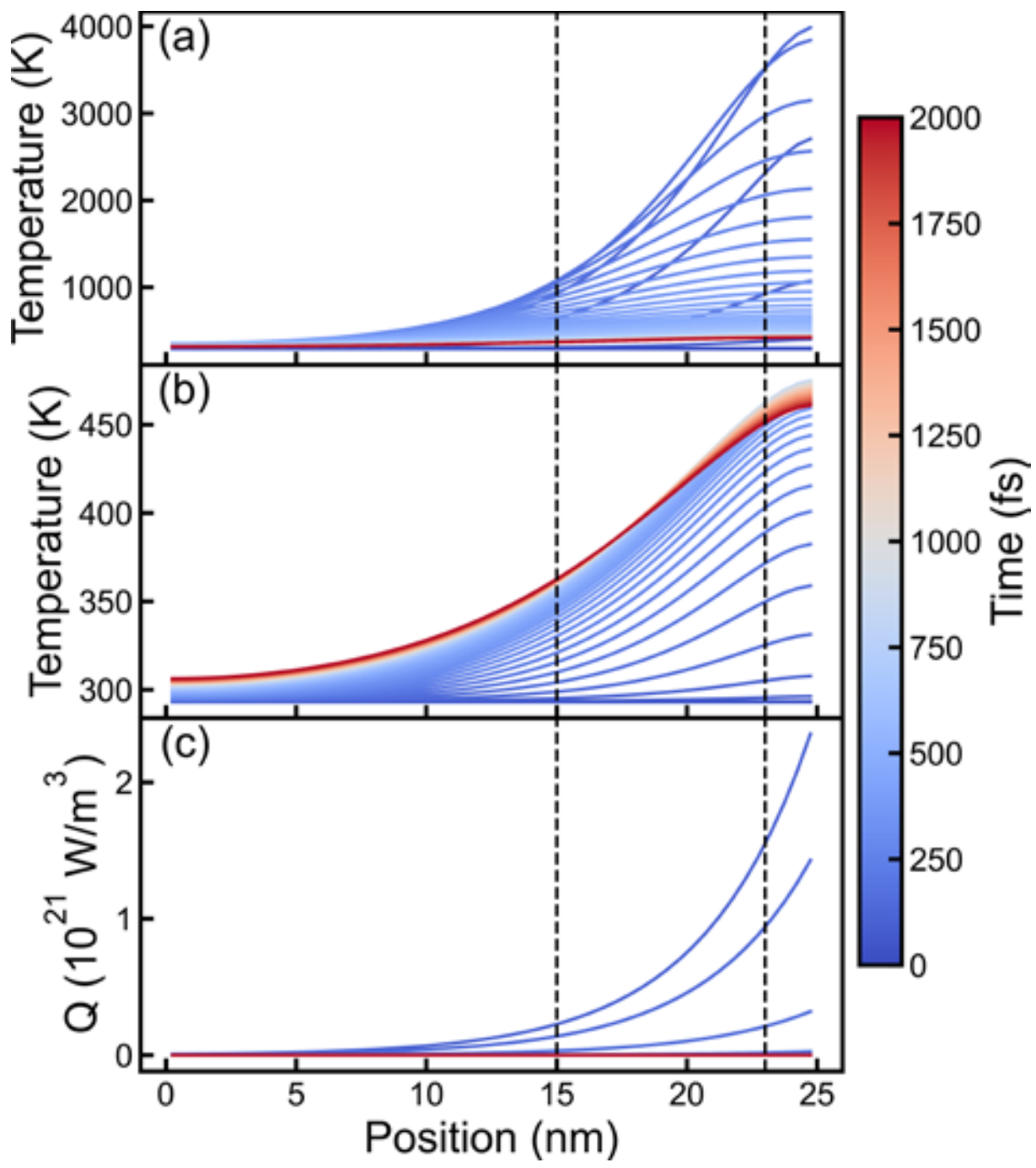

Figure S2 – Temperature and laser power density profiles as a function of $z$ position and time, corresponding to the results in Figure 1(a) in the main text with 0.1 ps pulse width. (a) Electron temperature, (b) lattice temperature, and (c) laser power density.

The laser pulse excites non-equilibrium electrons, proportional to the laser power density. Thus, initially, a gradient of non-equilibrium electron density, $n_s$, is generated as shown in Figure S3(b). These non-equilibrium electrons diffuse away from concentration gradients and, more importantly, superdiffuse away from electron temperature gradients. Thus, after the laser pulse, the non-equilibrium electron density becomes approximately uniform, and relaxes quickly as electrons recombine to lower energy levels. The displacement of these non-equilibrium electrons, under concentration and temperature gradients, results in generation of an excess charge density, as shown in Figure S3(a). As electrons are displaced from top to bottom, a positive excess charge density is generated at the top, and as electrons accumulate at the bottom of the spin valve, a negative excess charge density is generated there. This movement of non-equilibrium electrons also results in a positive charge current density, as shown in Figure S3(c). The current density is zero at the top and bottom surfaces, as expected. However, due to the change in charge density, the current density divergence is not zero. After the laser pulse, the generated excess charge density must relax. The relaxation process is diffusion of excess electrons away from accumulation gradients. This results in a negative current density, as shown in Figure S3(c). As diffusion is a much slower process compared to superdiffusion, the negative current density amplitude is smaller. Overall, the net charge displaced is zero, as expected from charge conservation. The overall charge current transient is thus composed of two phases: i) an initial positive current transient, generated due to superdiffusive forward flow of electrons, and ii) a negative current transient, generated due to diffusive backward flow of electrons.

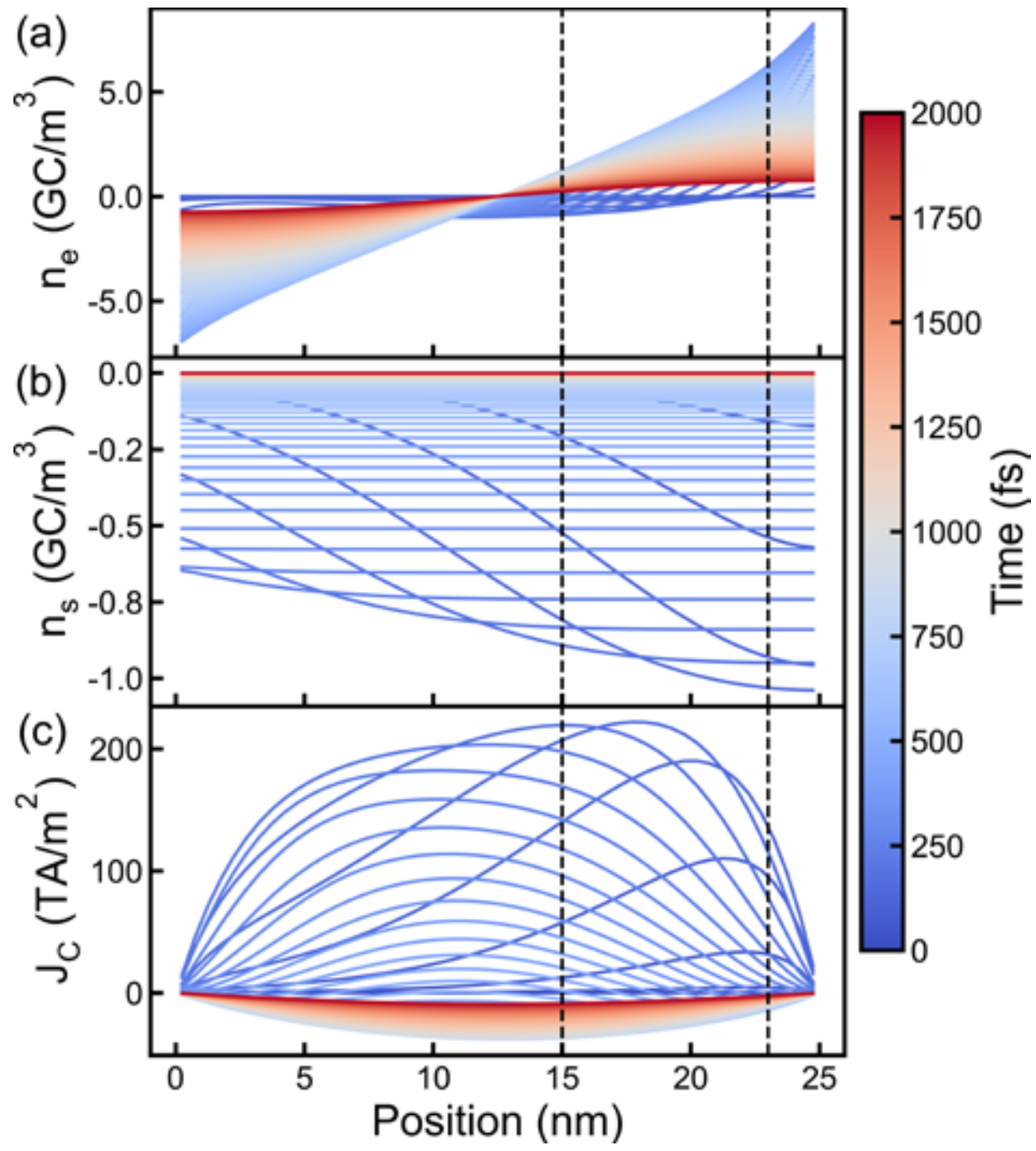

Figure S3 – Charge and current density profiles as a function of $z$ position and time, corresponding to the results in Figure 1(a) in the main text with 0.1 ps pulse width. (a) Excess charge density, (b) non-equilibrium charge density, and (c) charge current density.

The excess charge distribution also results in a long-ranged electrostatic interaction which can be included by solving Maxwell's first equation: $\nabla.\mathbf{E} = n_e / \varepsilon_0$. In particular, this results in the free current density contribution, $\mathbf{J}_f = \sigma\mathbf{E}$, where the electric field is due to separation of charges. The magnitude of this free current density is 2 – 3 orders of magnitude smaller than that due to constitutive relations for the problem studied here. For example consider Figure S3 showing the excess charge density and current density. An estimate of the free current density may be obtained from this by noting the excess charge density is approximately linear along the *z* axis position. With the boundary condition of zero electric field normal to external surfaces, the electric field is a square function of *z*, taking on the maximum magnitude of $n_{e,\max} a^2 / 8\varepsilon_0$, where *a* is the thickness of the spin valve. Using an extreme value of conductivity, $\sigma = 10^7$ S/m, the maximum free current density is $J_f = 0.6$ TA/m$^2$ for the maximum excess charge density shown in Figure S3. The corresponding maximum current density due to constitutive relations, shown in Figure S3, is ~200 TA/m$^2$.

## Spin-Dependent Seebeck Effect

The Seebeck effect, with a positive Seebeck coefficient, generates current transients similar to the non-equilibrium electron excitation mechanism. That is, a positive current transient is generated as electrons diffuse under temperature gradients from the hot to the cold side. As before, this process generates an excess charge density which must diffuse back due to accumulation gradients, resulting in a negative current transient. Examples are shown in Figure S4 for Seebeck coefficient $S_c$ = 10 μV/K, and conductivity $\sigma = 10^7$ S/m, which may be compared to the current transients shown in Figure 1 in the main text. These values are upper estimates for metallic layers, thus we expect the current transients due to the Seebeck effect are much smaller in practice. Even so, these current transients are an order of magnitude smaller than those generated through the low superdiffusive regime. Thus, the Seebeck effect is small in comparison with the superdiffusive mechanism in metallic layers.

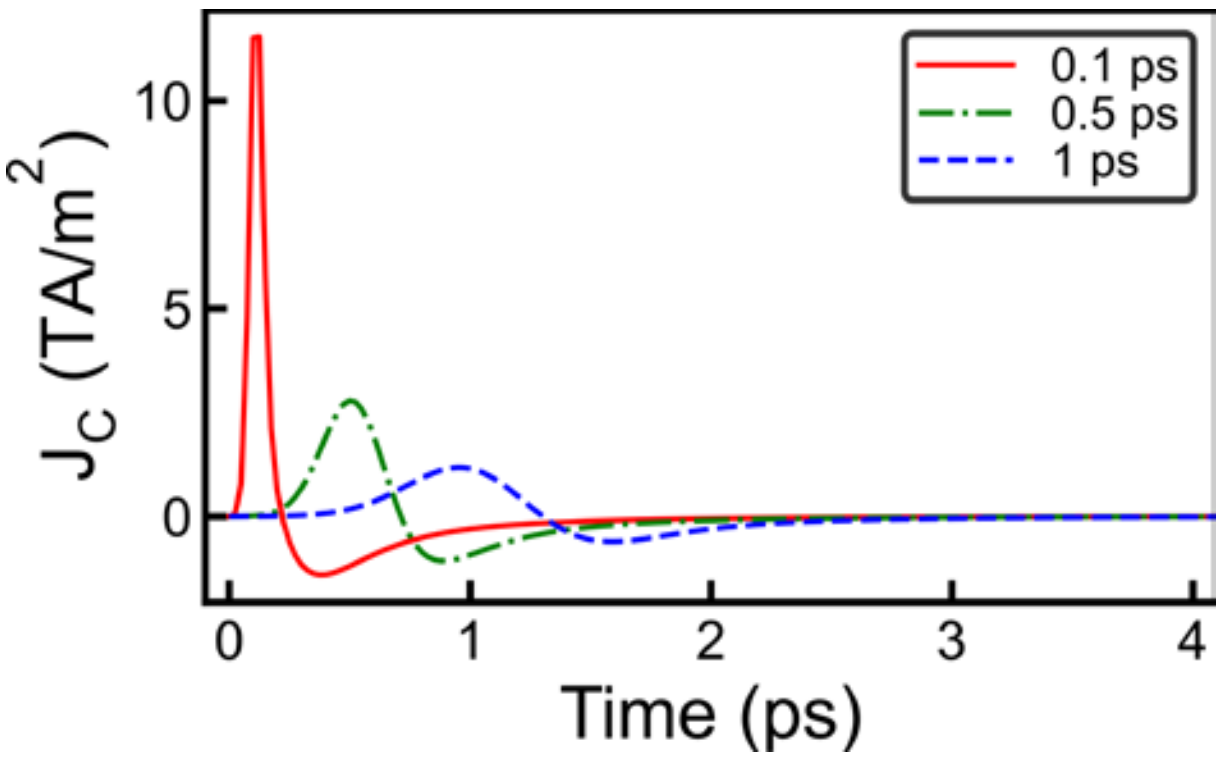

Figure S4 – Current transients generated through the Seebeck effect alone, at a fluence of 3 $mJ/cm^2$ and pulse widths of 0.1 ps, 0.5 ps, and 1 ps, for the same spin valve structure as that in Figure 1 in the main text.

## Analysis of spin torque

The spin torque generated in the spin valve structure due to the transient flow of electrons is similar to the Slonczewski STT, $\mathbf{T}_S = \eta(\mu_B / e)(J_C / d_{FL})\mathbf{m}\times(\mathbf{m}\times\mathbf{p})$. Here **m** and **p** are the magnetization directions in the FL and RL, respectively, and $\eta$ is the STT polarization parameter. The self-consistent spin torque is computed using the ASDD model, and the Slonczewski STT is computed using the above equation, for the same charge current density transient. The laser fluence generating the current transient is 0.1 mJ/cm$^2$, which does not induce switching. Results are shown in Figure S5 for the P+ state. For longer pulses, 2 ps and 5 ps, it is seen the self-consistent and Slonczewski spin torques are similar. The main difference is the self-consistent spin torque lags the Slonczewski STT, by approximately 0.5 ps. This is due to the spin accumulation response time, as a changing charge current density transient does not result in an instantaneous change in **S** – for the Slonczewski STT this dynamical effect is ignored. Thus, in the limit of long pulse width, the two spin torques are essentially the same. However, as the pulse width is reduced below 1 ps, becoming comparable to the spin accumulation response time, the two spin torques become significantly different, as seen in Figure S5. This shows the Slonczewski STT is not sufficiently accurate in the ultrafast regime, and it is necessary to take into account the spin accumulation response time.

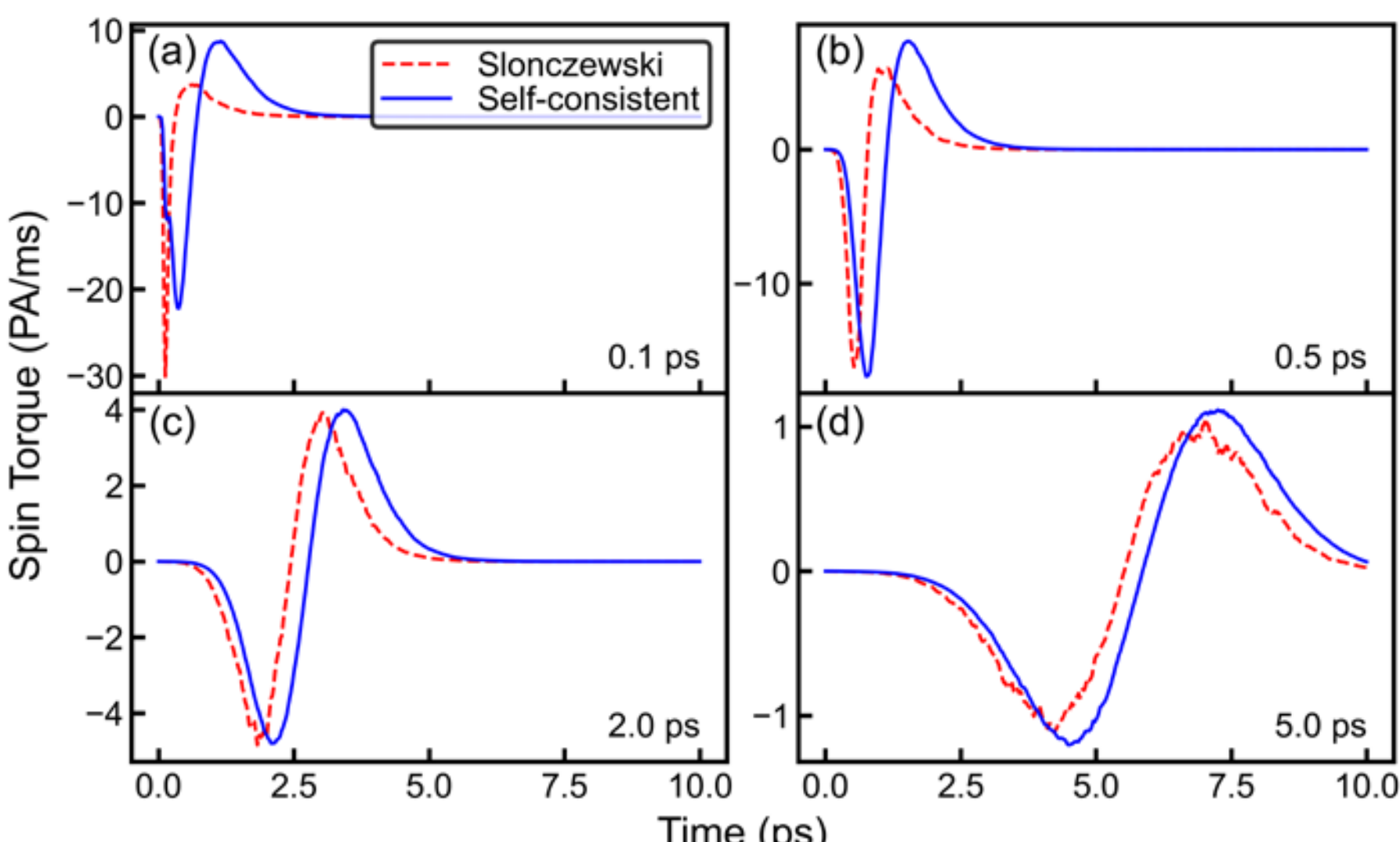

Figure S5 – Computed spin torques in the FL for 0.1 mJ/cm$^2$ fluence and high superdiffusive regime with no spin pumping, using $d_{RL}$ = 15 nm and $d_{SL}$ = 0.8$\lambda_{sf}$. Slonczewski and self-

consistent spin torques are shown, as indicated, for (a) 0.1 ps, (b), 0.5 ps, (c) 2.0 ps, and (d) 5.0 ps pulse widths.

Further details are shown in Figure S6, where the negative and positive spin torque amplitudes are plotted as a function of pulse width. As seen, the amplitudes for the two spin torques are the same above 1 ps, but diverge as the pulse width is reduced.

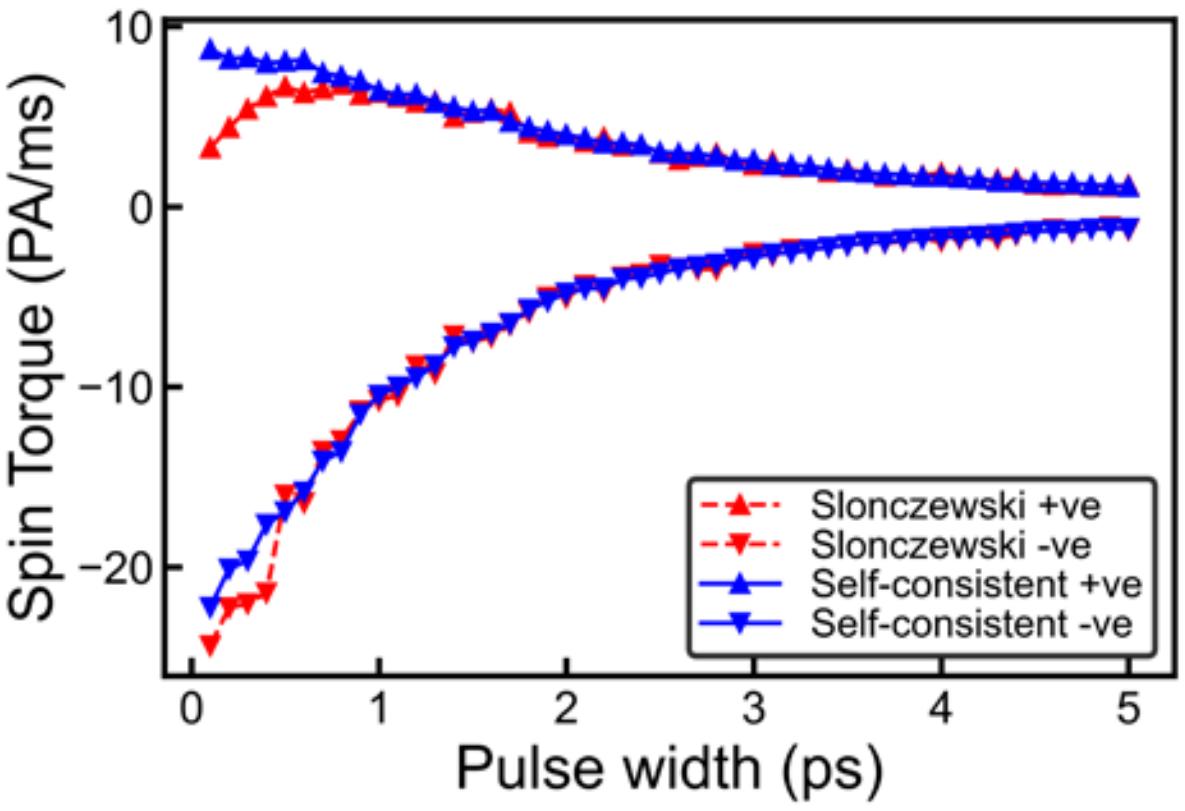

Figure S6 – Spin torque amplitudes for the self-consistent and Slonczewski spin torques, plotted as a function of pulse width. These are extracted from the computed spin torques, as shown in Figure S5, for the negative and positive amplitudes.

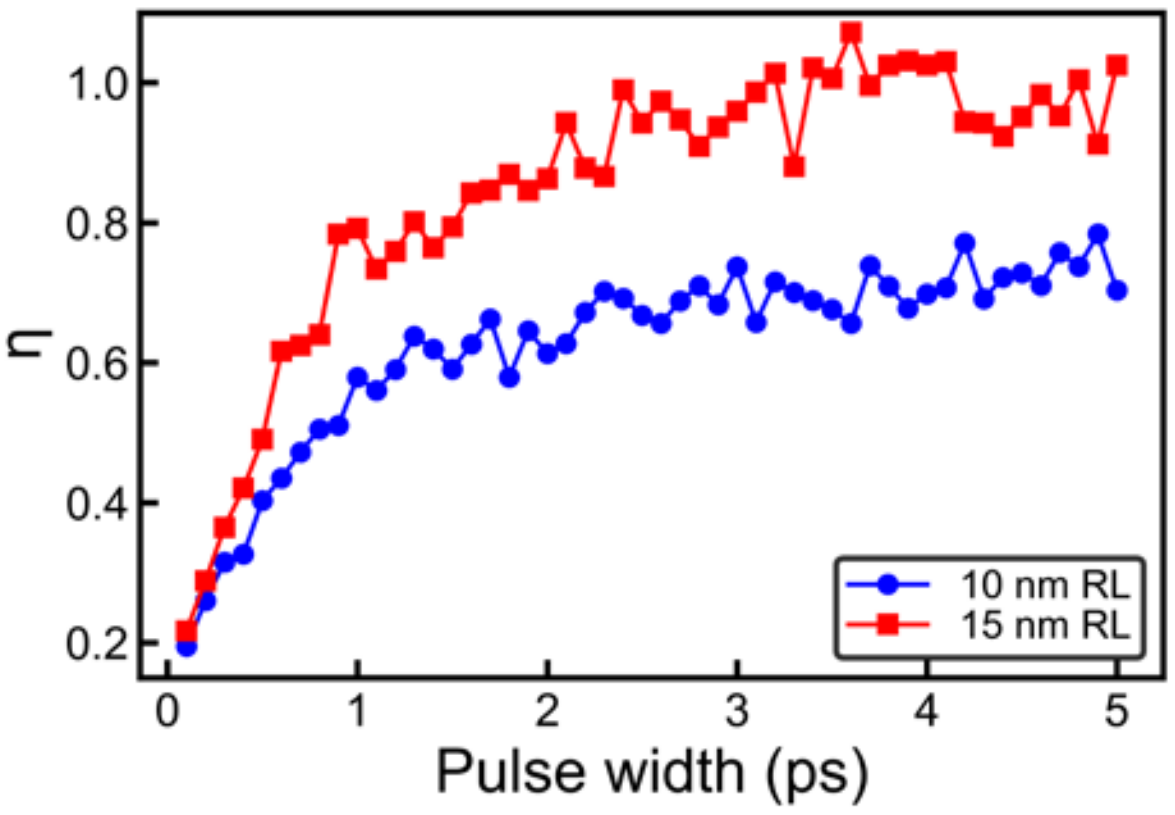

Figure S7 – STT polarization, η, as a function of pulse width, computed by fitting the Slonczewski torque to the self-consistent torque. The calculation is repeated for two different thicknesses of RL, 10 and 15 nm.

The STT polarization is used as a fitting factor here, adjusted to give the same Slonczewski STT amplitudes as the self-consistent spin torque. This is shown in Figure S7.

Finally, the temperature dependences of the spin torque amplitudes are shown in Figure S8, where the heat solver is disabled and temperatures in the FL and RL are fixed for each laser pulse. In particular, increasing the RL temperature decreases the spin torque amplitudes, as seen in Figure S8(b) for FL temperature of 300 K. This is due to the lower average spin polarization exerted by the RL on the FL, as expected. On the other hand, increasing the FL temperature, for a fixed RL temperature of 300 K, results in increasing spin torque amplitudes up to $T_C$, and these are seen to plateau for $T > T_C$, as shown in Figure S8(a). This is due to the larger angle between **m** and **p**, on average, resulting from increased thermal fluctuations.

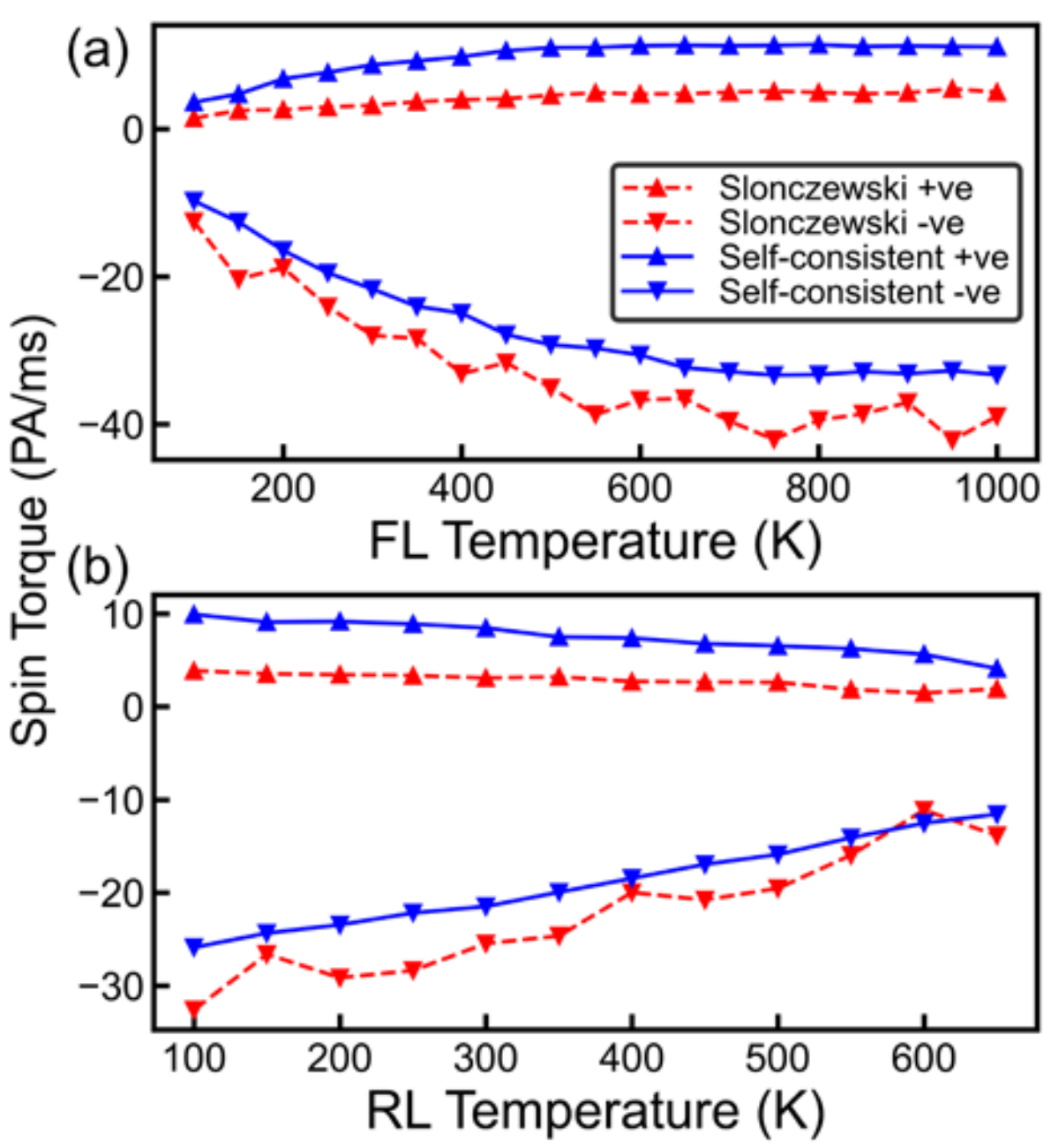

Figure S8 – Spin torque amplitudes as a function of temperature, for 0.1 ps pulse width and 0.1 mJ/cm$^2$ fluence. Here, the heat solver is disabled and the temperature is fixed throughout each laser pulse. (a) Fixed RL temperature of 300 K, with spin torque amplitudes plotted as a function of FL temperature. (b) Fixed FL temperature of 300 K, with spin torque amplitudes plotted as a function of RL temperature.

## Magnetization switching details

Example P+ to AP+ (3 mJ/cm$^2$ fluence and 0.1 ps pulse width) and AP+ to P+ (5 mJ/cm$^2$ fluence and 1 ps pulse width) switching events are shown in Figure S9, plotting the magnetization in the FL for $d_{RL}$ = 15 nm and $d_{SL}$ = 0.8$\lambda_{sf}$. The switching occurs during the initial phase, on a timescale comparable to the pulse width, characterized by crossing the *x* axis as the magnetization changes sign. This is followed by a longer recovery, on a timescale of 100 ps or more, as the temperature gradually reduces back to room temperature.

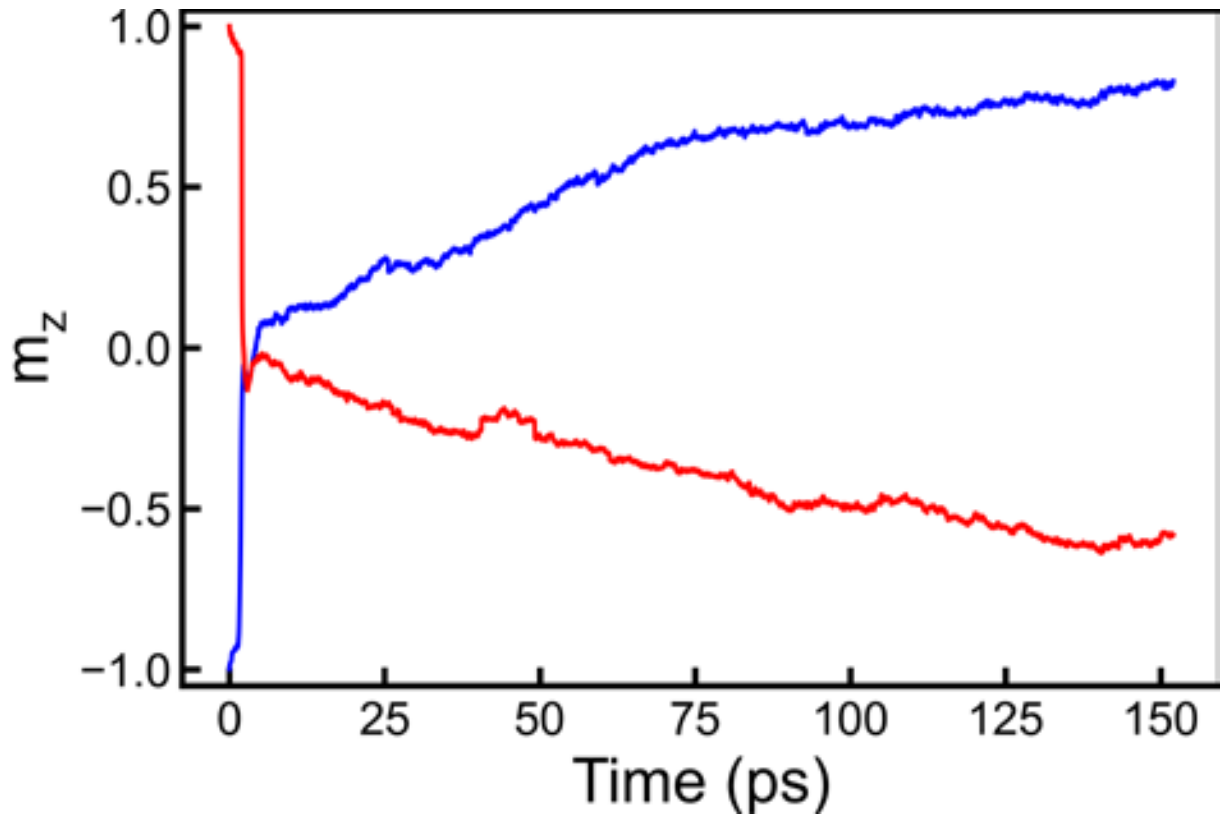

Figure S9 – FL magnetization switching dynamics in the low superdiffusive regime, with $d_{RL}$ = 15 nm, $d_{SL}$ = 0.8$\lambda_{sf}$. The red line shows P+ to AP+ switching with 3 mJ/cm$^2$ fluence and 0.1 ps pulse width, and the blue line shows AP+ to P+ switching with 5 mJ/cm$^2$ fluence and 1 ps pulse width.

Details are shown in Figure S10 for starting P+ state and different fluences. For lower fluences (1.5 mJ/cm$^2$ here), the magnetization does not cross the *x* axis, and consequently recovers back to the P+ state. For fluences above the threshold value (3 mJ/cm$^2$ here), the magnetization crosses the *x* axis and recovers to the AP+ state. This is similar to the magnetization dynamics shown in [Nat. Mater. 22, 725 (2023)]. For even larger fluences (6 mJ/cm$^2$ here), where AP+ to P+ switching is possible, the magnetization initially crosses the *x* axis, but is followed by crossing back the *x* axis and recovery to the P+ state. The initial *x* axis crossing is due to a combination of the forward flow of electrons and accumulation of minority spins at the RL, as well as sufficient demagnetization in the FL. Re-crossing of the *x* axis occurs at larger fluences where the diffusive backward flow of electrons, repolarized by the RL, is sufficiently strong.

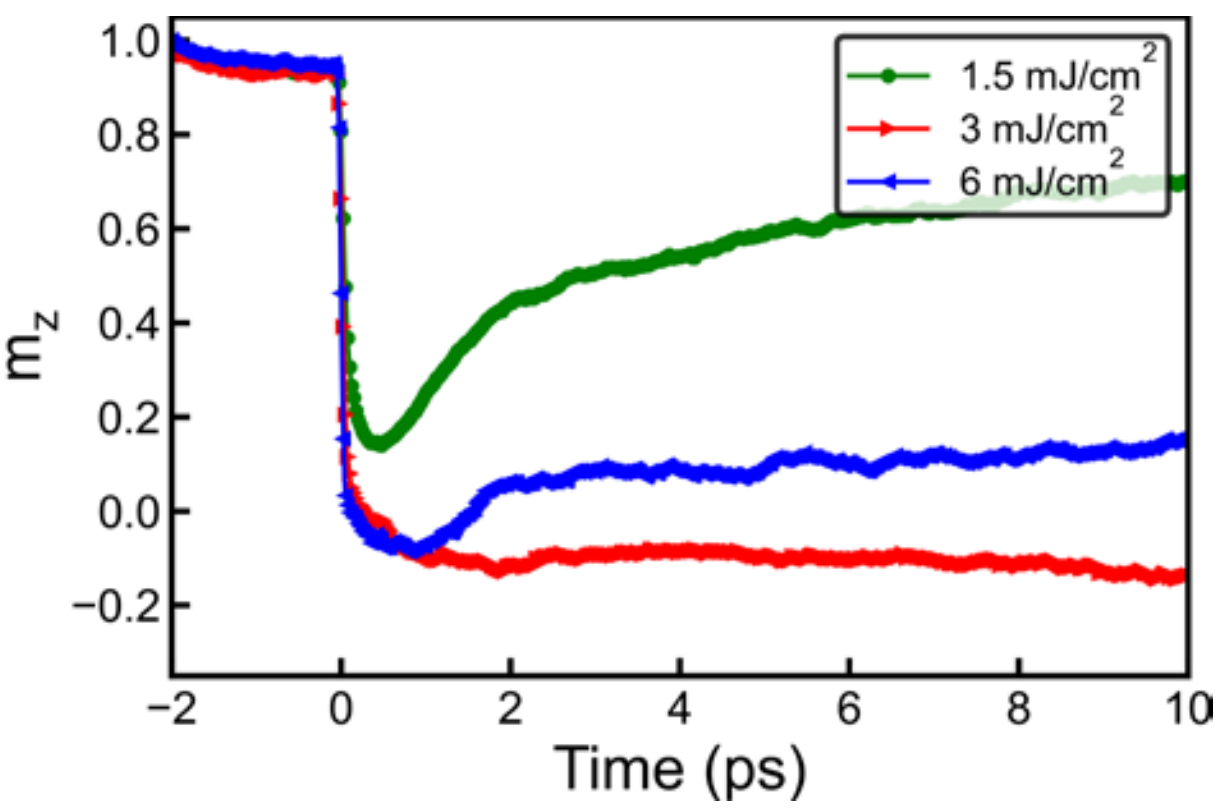

Figure S10 – Details of FL magnetization switching dynamics in the low superdiffusive regime, with $d_{RL}$ = 15 nm, $d_{SL} = 0.8\lambda_{sf}$, and 0.1 ps pulse width, starting from the P+ state. The different lines show the results obtained with different fluences, as indicated in the legend.

Further details are shown next, through time and spatially resolved magnetization switching profiles. The P+ to AP+ switching event is plotted in Figure S11, corresponding to that shown in Figure S10 for 3 mJ/cm$^2$ fluence, showing the $m_z$ normalized magnetization component plotted through the thickness of the spin valve, thus including both FL and RL. In the FL a rapid demagnetization process occurs, followed by a longer magnetization recovery process, as discussed above. The RL is also affected, particularly in the part closer to the FL. The decrease in magnetization here is largely due to the temperature increase, which decays exponentially away from the laser-exposed FL surface, resulting in stronger magnetization decrease closer to the FL. Similar observations apply to the AP+ to P+ switching event shown in Figure S12.

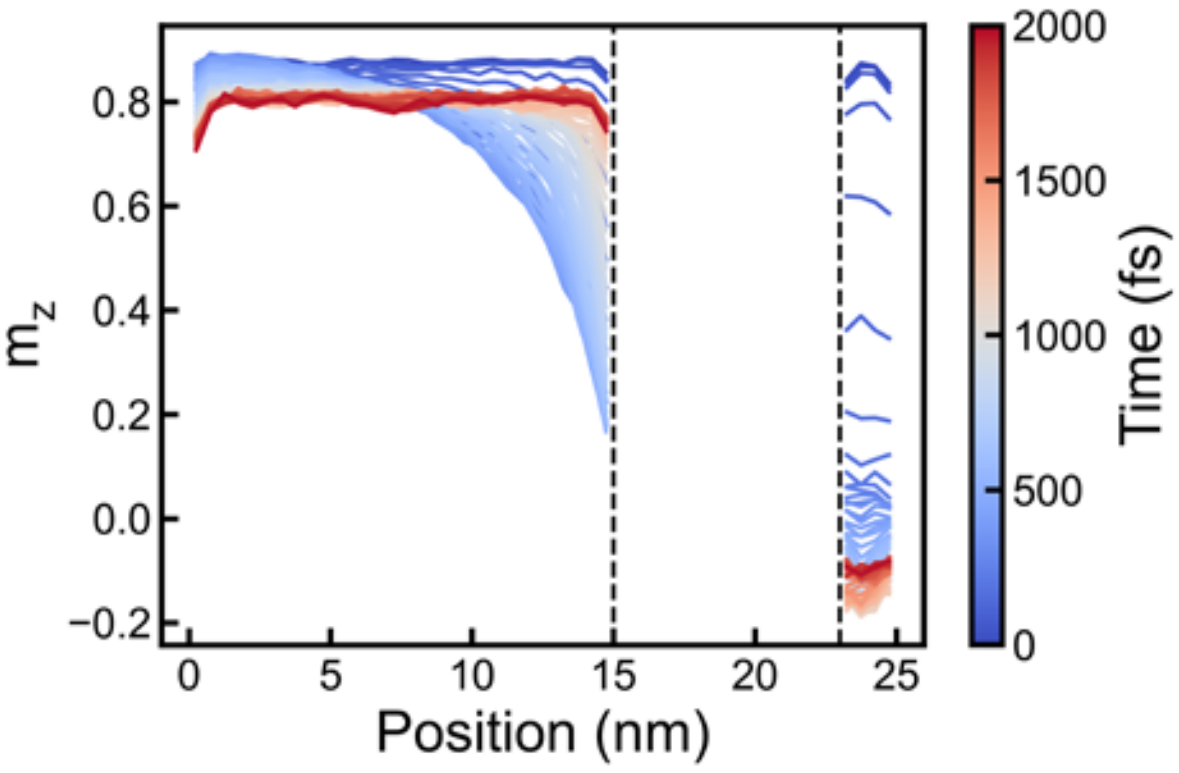

Figure S11 – P+ to AP+ magnetization switching profiles as a function of time. The fluence is 3 mJ/cm$^2$ with 0.1 ps pulse width, where $d_{RL}$ = 15 nm and $d_{SL} = 0.8\lambda_{sf}$, in the low superdiffusive regime.

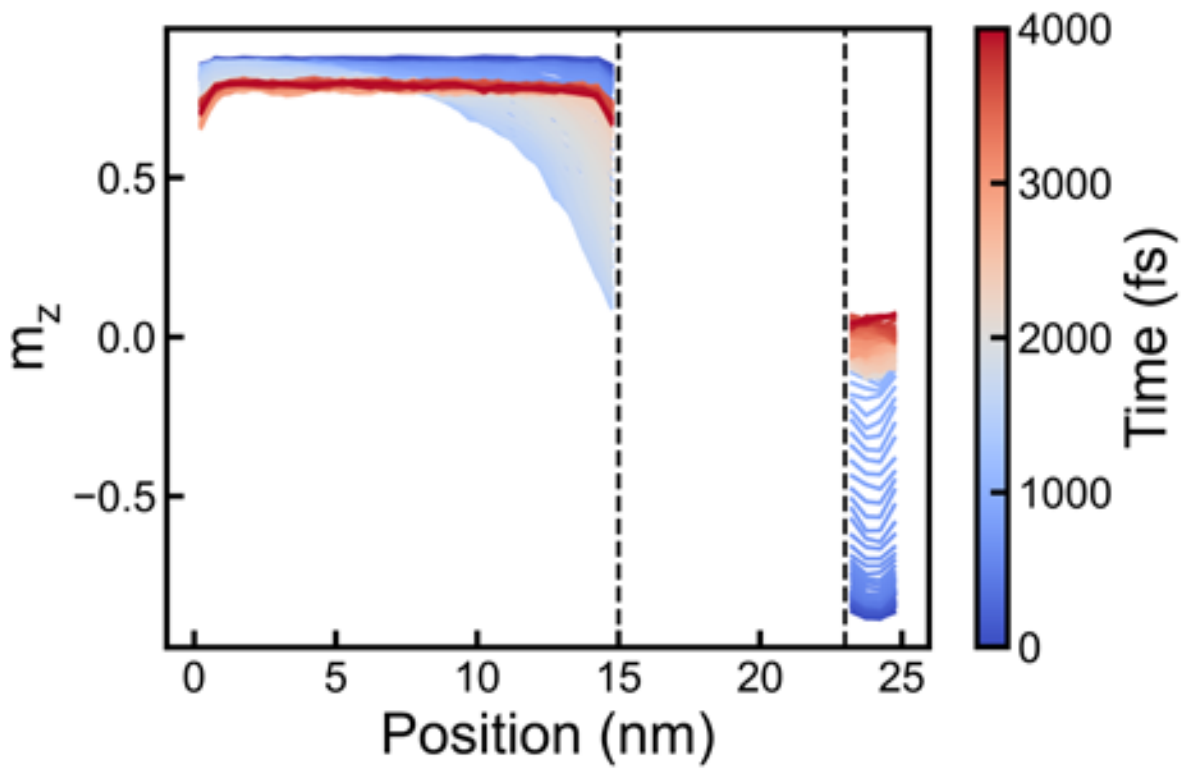

Figure S12 – AP+ to P+ magnetization switching profiles as a function of time. The fluence is 5 mJ/cm$^2$ with 1 ps pulse width, where $d_{RL}$ = 15 nm and $d_{SL}$ = 0.8$\lambda_{sf}$, in the low superdiffusive regime.

## Pure spin currents and switching due to spin pumping

The effect of pure spin currents due to spin pumping on magnetization switching, in the absence of the superdiffusive electron mechanism, is discussed in this section. Representative spin current transients are shown in Figure S13.

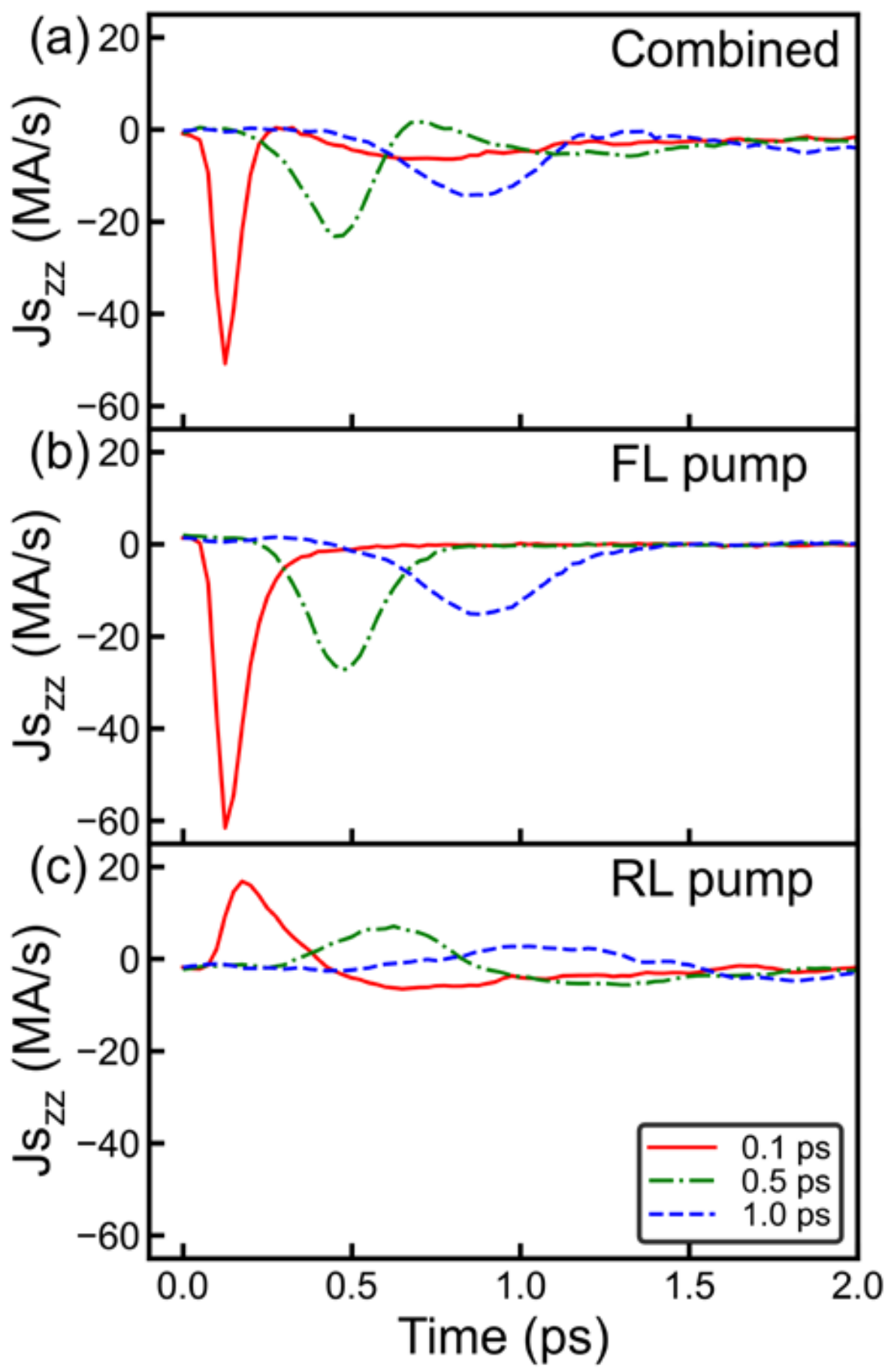

Figure S13 – Laser-induced pure spin current transients due to spin pumping, using $d_{RL}$ = 15 nm and $d_{SL}$ = 0.8$\lambda_{sf}$, for fluence 3 mJ/cm$^2$ and varying pulse width. Spin currents in SL generated by (a) both RL and FL, (b) FL only, and (c) RL only.

As discussed in the main text, starting from the P+ state the RL pumps a positive spin current in the SL during demagnetization, as required by conservation of total angular momentum. This is followed by a negative spin current transient as the magnetization recovers. The spin current transients pumped by the RL in the SL, for pulse widths of 0.1 ps, 0.5 ps and 1 ps are shown again in Figure S13(c). The FL also pumps a spin current into the SL, and in particular pronounced spin current transients are observed during demagnetization, as plotted in Figure S13(b). The sign of this transient is opposite to that pumped by the RL since the FL/SL stacking is inverse to that of SL/RL. The overall spin current transients, which are sums of the above two contributions, are shown in Figure S13(a).

Pure spin current transients also give rise to spin torques, although as we show below this effect is weaker compared to the superdiffusive mechanism and, as discussed in the main text, does not explain the experimental results shown in [Nat. Mater. 22, 725 (2023)]. In particular, the pure spin current pumped during the magnetization recovery process favors an antiparallel alignment of RL and FL, whilst the spin current pumped during demagnetization has no effect on the final magnetization configuration – this is in agreement with the analysis in [Nat. Mater. 22, 725 (2023)]. Switching probabilities using only the spin pumping mechanism, are calculated for the same spin valve used in the main text, with $d_{RL}$ = 15 nm and two different SL thicknesses, 2.5 nm and 20 nm (corresponding to $d_{SL} = 0.25\lambda_{sf}$ and $2.00\lambda_{sf}$ respectively). The switching probabilities are calculated as a function of fluence and pulse width, starting from states P+ as well as AP+. Results are shown in Figure S14.

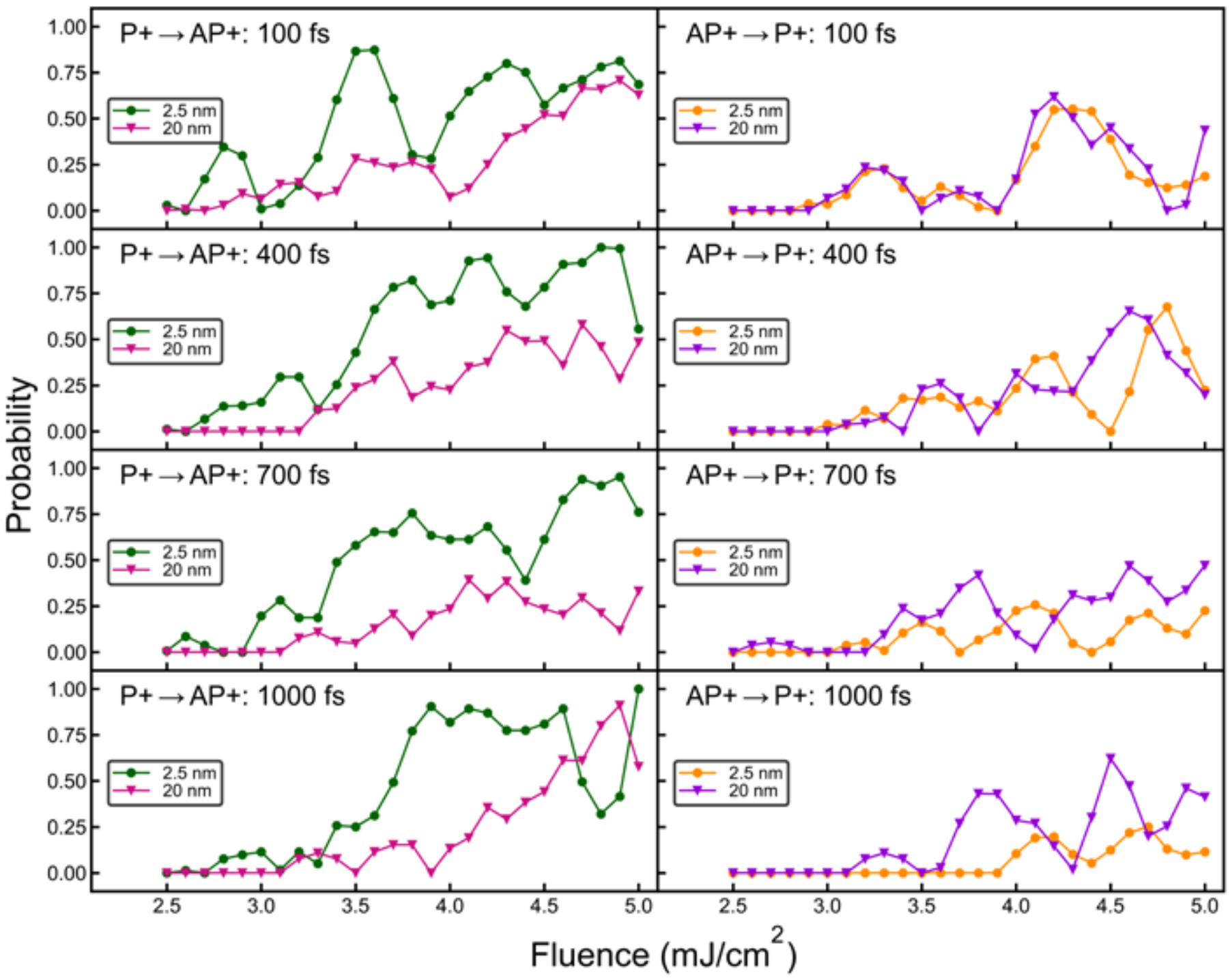

Figure S14 – Calculated switching probabilities using pure spin currents only for very high spin pumping regime ($\chi_{sp}$ = 36 kA/m). P+ and AP+ starting states are used for a range of pulse widths ranging from 100 fs to 1000 fs as indicated. The calculations are repeated for SL thicknesses of 2.5 nm and 20 nm.

A stronger spin pumping regime is used than in the main paper ($\chi_{sp}$ = 36 kA/m) in order to emphasize this effect. Since the spin currents decay in the SL, two thickness values are chosen, namely 2.5 nm and 20 nm, where the effect of spin pumping should be negligible in the latter. Starting from the AP+ state, increasing the fluence results in increased switching probabilities above ~3 mJ/cm$^2$, reaching an average maximum of 0.5. This is the same behavior obtained through purely thermal effects. When the FL is completely demagnetized above a threshold fluence which brings the maximum electron temperature above $T_C$, subsequent magnetization recovery is equally likely in either the P+ or AP+ configuration. Switching can occur, but is not deterministic. The threshold fluence also depends on the pulse width, since the maximum electron temperature decreases as the pulse width is increased, for the same fluence. This effect is seen in Figure S14. Moreover, no significant difference is observed when starting from the AP+ state for the two extreme SL thickness values. These results confirm that the effect of spin pumping on AP+ to P+ switching is negligible. On the other hand, when starting from the P+ state a consistent difference between the two SL thickness values are observed. Again, with an SL thickness of 20 nm, where the effect of spin pumping is negligible, the switching probabilities are similar to those obtained through thermal effects alone. With an SL thickness of 2.5 nm the switching probabilities are consistently larger above the threshold value, even reaching the maximum value of 1.0. These results show that spin pumping can generate P+ to AP+ switching, but not AP+ to P+ switching. However, the effect is much weaker than the superdiffusive mechanism.

## Effect of laser pulses on FM/NM bilayers

In simple FM/NM bilayers deterministic switching is not achieved, as expected. Instead, increasing the laser fluence increases the switching probability up to an average value of 0.5, corresponding to thermally activated random switching during the recovery phase. This is shown in Figure S15, comparing switching probabilities in a FM(2 nm)/NM(18 nm) bilayer for cases with self-consistent STT and no STT. For the no STT case the switching probability reaches a maximum average value of 0.5, as expected for purely random thermally activated switching. The STT case is similar, however the switching probabilities are consistently lower for the shorter pulses. This is likely due to a self-torque effect in the FM, resulting in a slight suppression of demagnetization during the superdiffusive forward flow current transient. As superdiffusive electrons originate at the top surface of the FM they carry angular momentum towards the bottom surface. Since the FM starts in a uniformly magnetized state, this transfer of angular momentum serves to reinforce it, impeding larger angle differences between magnetic moments through the thickness of the FM. This effect is more pronounced for shorter pulses, where the forward flow transients are stronger, with the switching probabilities tending towards those observed with purely thermal effects for pulses of 1 ps duration and longer.

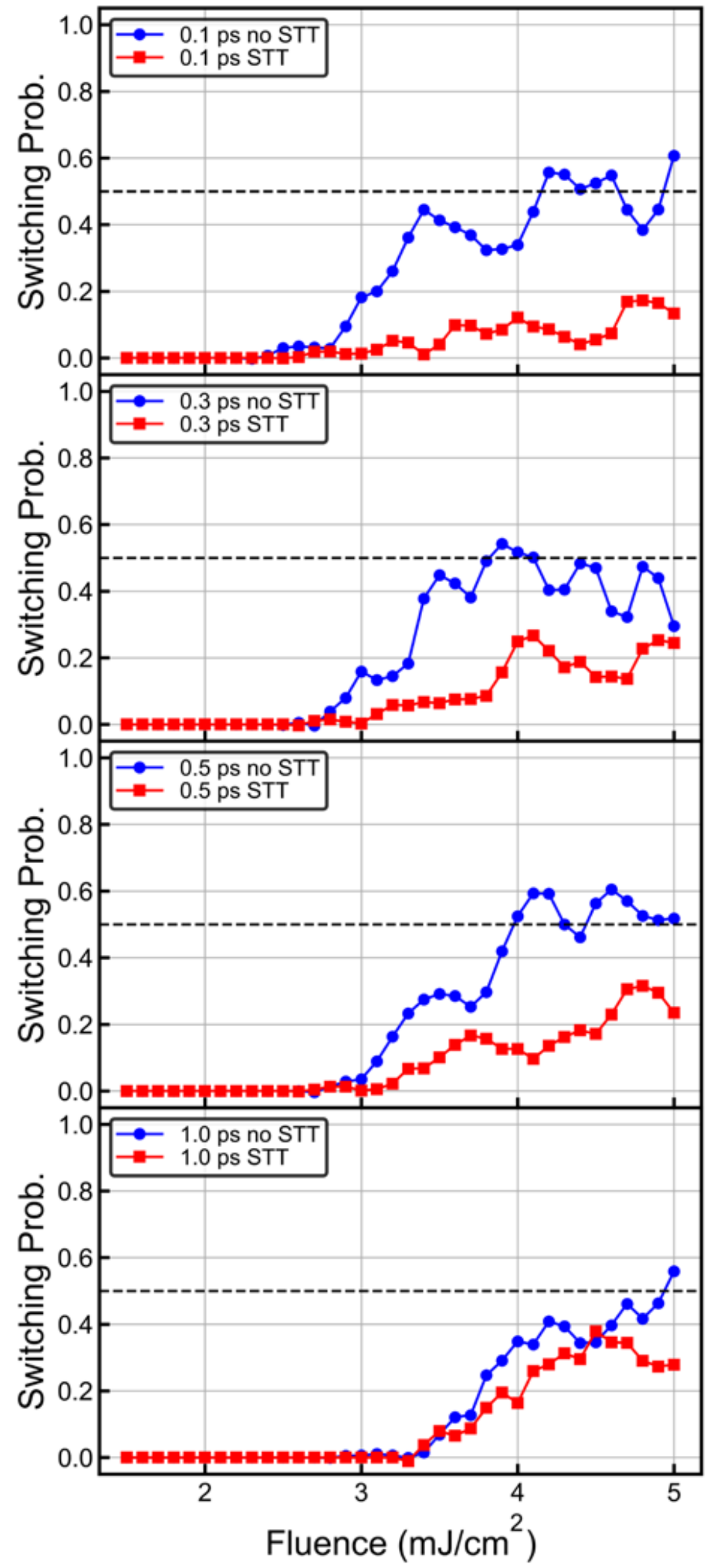

Figure S15 – Effect of laser pulses of varying fluence and pulse widths of 0.1 ps, 0.3 ps, 0.5 ps, and 1.0 ps, shown from top to bottom, respectively, on the switching probability of FM(2 nm)/NM(18 nm) bilayers. The switching probabilities were calculated with the self-consistent spin torque in the high superdiffusive regime, as well as without. The expected maximum average switching probability of 0.5, obtained when the fluence is large enough to sufficiently demagnetize the FM layer, is indicated using the dashed black lines.

AOS with a Gaussian laser spot

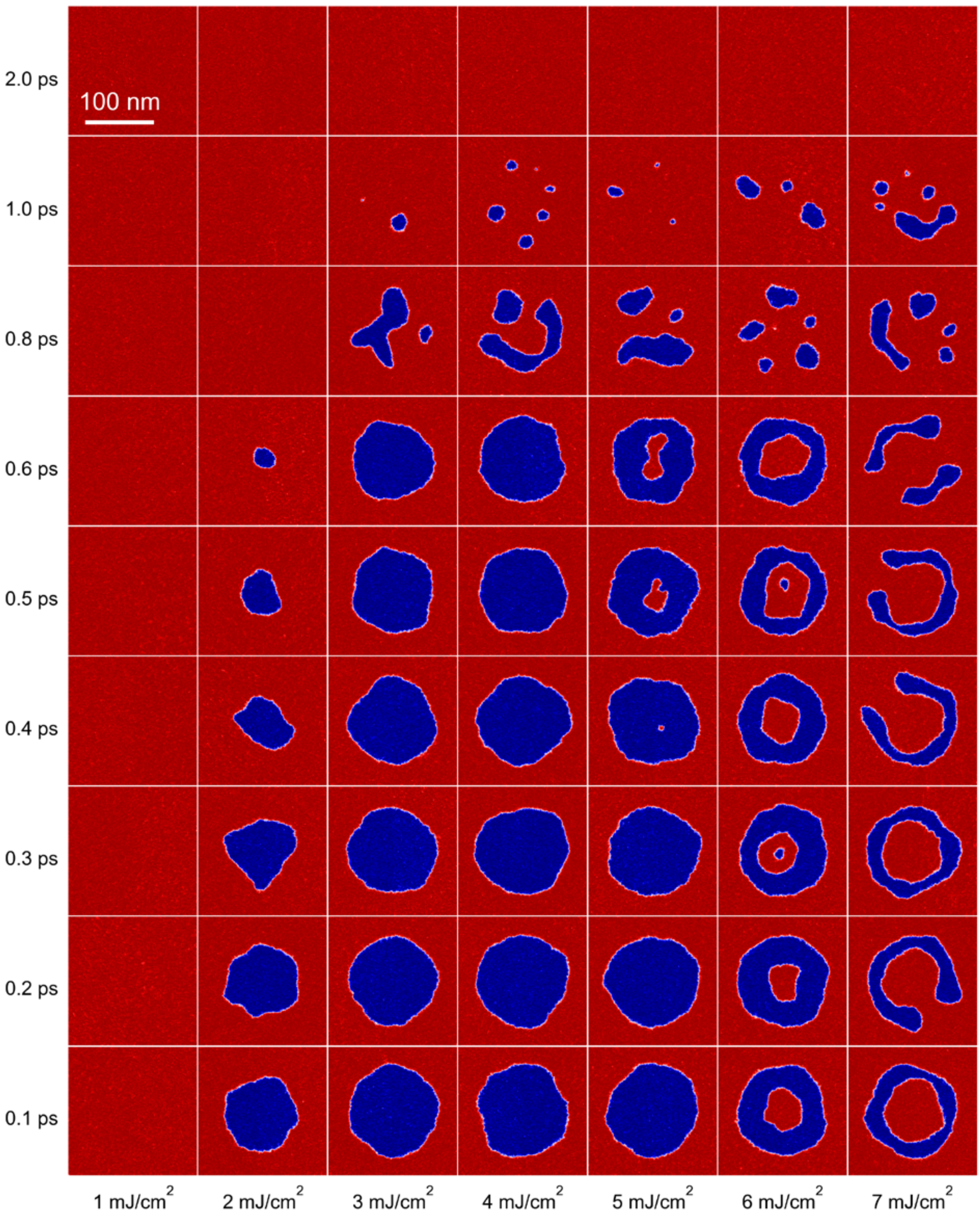

Figure S16 – Switching map using a circular laser spot with Gaussian profile, as a function of pulse width and fluence. The starting state was P+, and the images show the resulting states in the FL after 150 ps. Red is P+, and blue is AP+.

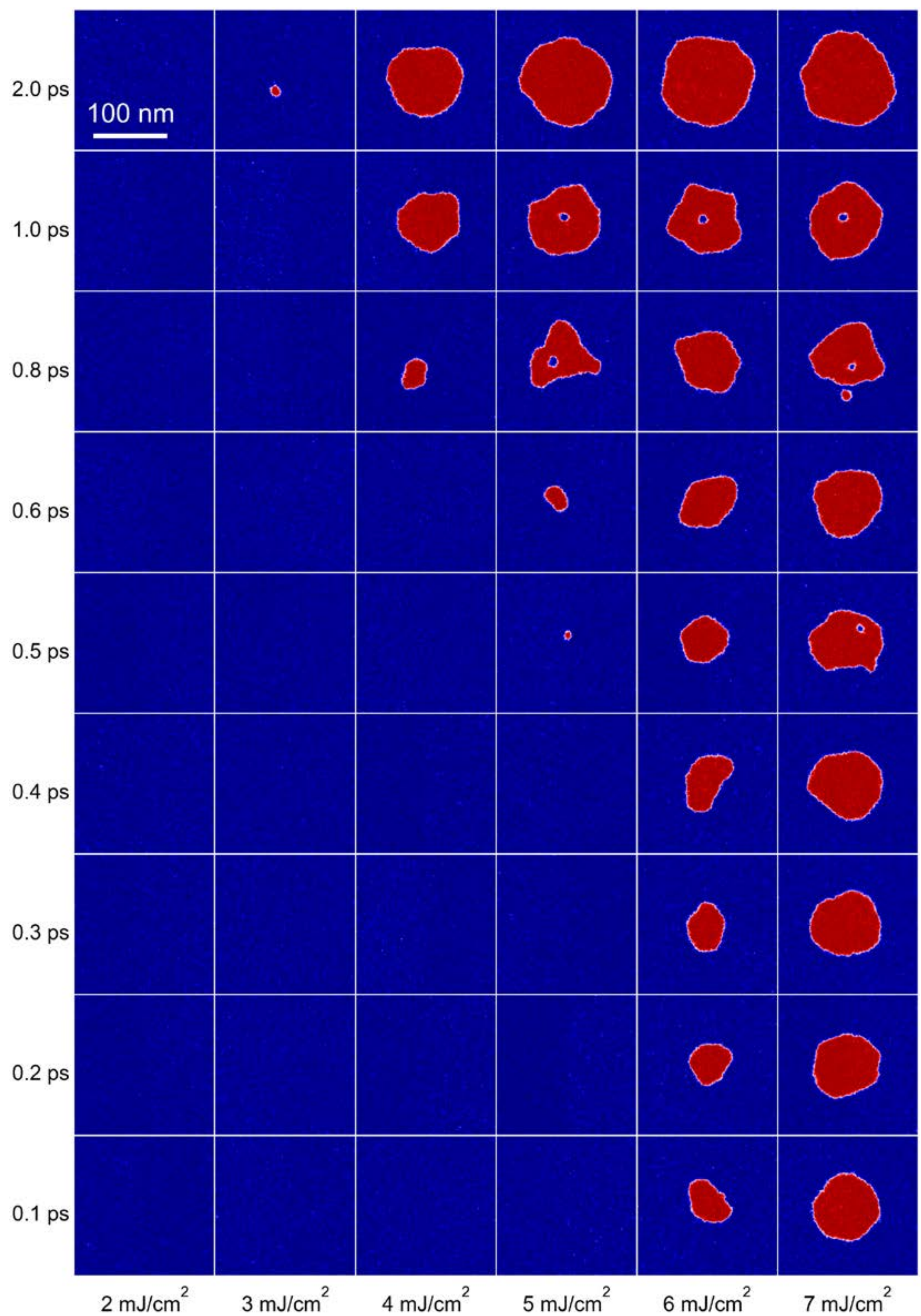

Figure S17 – Switching map using a circular laser spot with Gaussian profile, as a function of pulse width and fluence. The starting state was AP+, and the images show the resulting states in the FL after 150 ps. Red is P+, and blue is AP+.

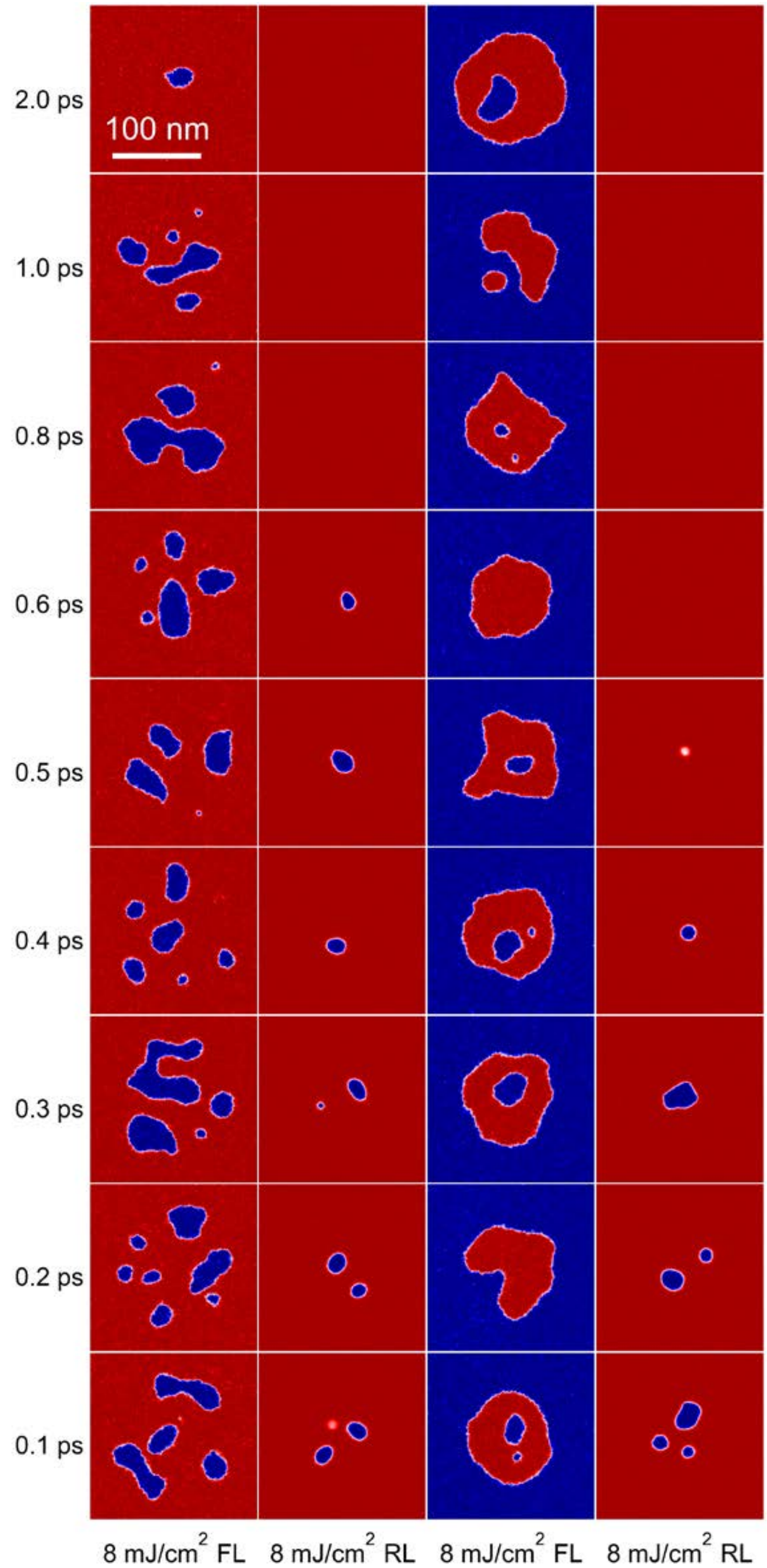

Figure S18 – Switching map using a circular laser spot with Gaussian profile, as a function of pulse width at high fluence of 8 mJ/cm$^2$. The starting states were P+ and AP+ for the first and last 2 columns respectively. The images show the resulting states in the FL and RL, as indicated by the labels at the bottom, after 150 ps.

Results obtained with a circular laser spot with Gaussian intensity profile are shown in Figure S16 and Figure S17 for P+ and AP+ starting states respectively, and also in Figure S18 for the largest simulated fluence of 8 mJ/cm$^2$. Both the fluence and pulse width are varied, the latter between 0.1 ps and 2.0 ps. These results, obtained in the high superdiffusive regime, substantiate those shown in Figure 4 in the main text. For reference, the maximum electron and lattice temperatures as a function of laser fluence and pulse width are shown in Figure S19.

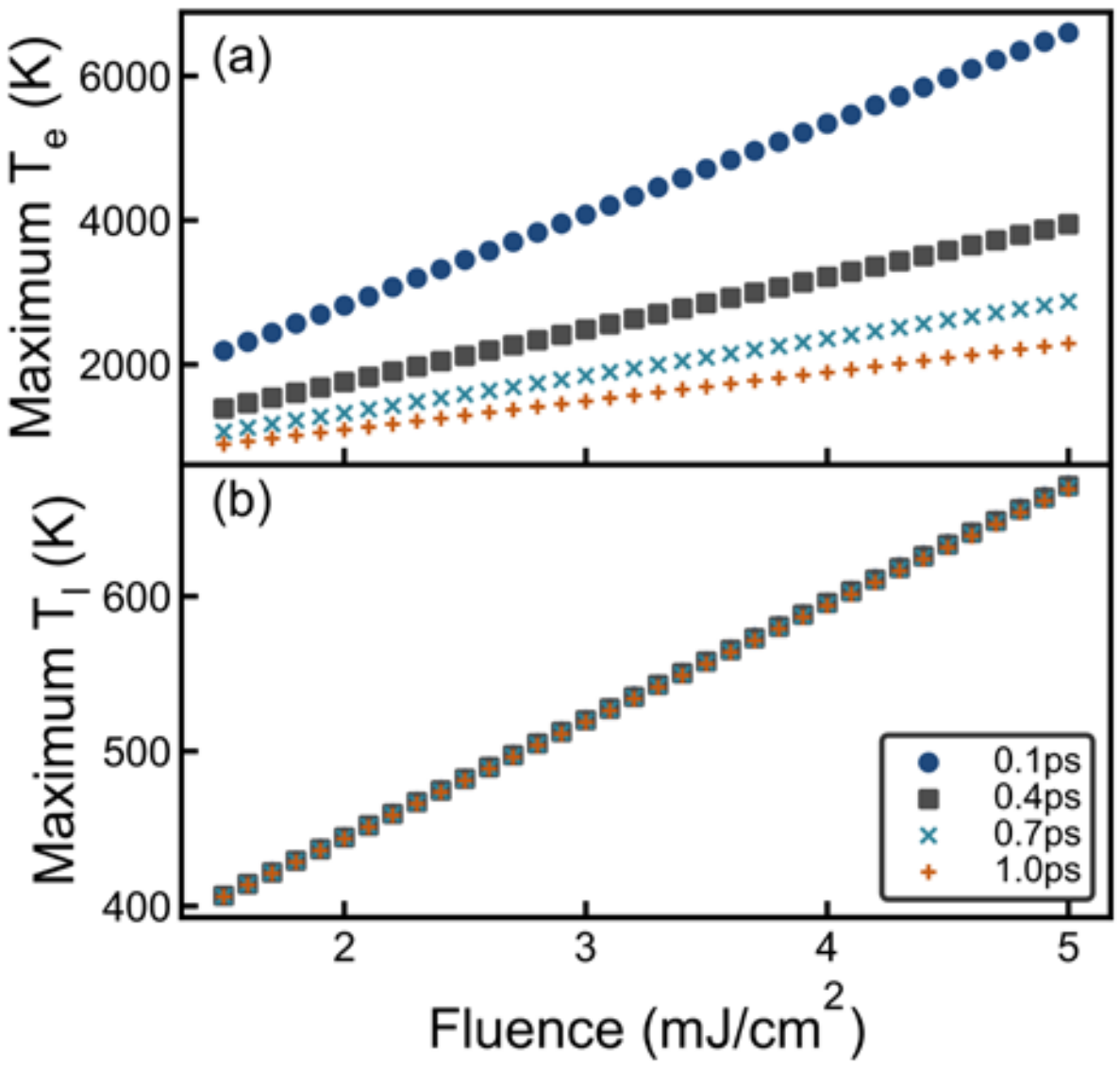

Figure S19 – Maximum temperatures in the FL as a function of laser fluence, shown for different pulse widths, for (a) electron temperature, and (b) lattice temperature.

First, switching starting from the P+ configuration is discussed, with results shown in Figure S16. At a low threshold fluence P+ to AP+ switching is obtained (2 mJ/cm$^2$). This occurs on the superdiffusive positive current transient. As the pulse width is increased the amplitude of the positive current transient decreases, resulting in a decrease of the switched area, and eventually no switching is obtained – see the 2 mJ/cm$^2$ fluence series. Increasing the fluence increases both the positive and negative current transient amplitudes. This allows full P+ to AP+ switching under the laser spot for longer pulses, however multi-domain structures are obtained by further increasing the pulse width – see the 3 and 4 mJ/cm$^2$ fluence series. The multi-domain structures are obtained initially due to incomplete switching to the AP+ configuration. Increasing the pulse width results in decrease of the positive current transient amplitude, to the point where fully deterministic switching under the laser spot is not possible. Further increasing the pulse width results in a P+ final state as the positive current transient is too small to affect the starting state. As the fluence is increased the negative current transient becomes sufficiently strong to affect switching. This results in a type of multi-domain structure where an AP+ configuration is obtained under the laser spot during the positive current transient, however during the negative current transient a smaller P+ region is formed inside the larger AP+ region – see the 5 mJ/cm$^2$ fluence series in Figure S16, as well as Figure 4 and related discussion in the main text. This effect becomes more pronounced when increasing the fluence, with the P+ region growing in size – see the 6 mJ/cm$^2$ fluence series in Figure S16. Further increasing the fluence results in a merging of the AP+ and P+ region, which leads to complex multi-domain structures – see the 7 mJ/cm$^2$ fluence series, as well as the longer pulse widths in the 5 and 6 mJ/cm$^2$ fluence series in Figure S16. Thus it is seen the pulse width threshold for multi-domain formation decreases as the fluence is increased.

Switching starting from the AP+ configuration is discussed next, with results shown in Figure S17. AP+ to P+ switching is governed by the negative current transient, and as discussed in the main text a larger fluence threshold is required compared to P+ to AP+ switching. Increasing the pulse width results in full switching to the AP+ configuration under the laser spot as the negative current transient, and associated spin torque, acts for a longer duration. Multi-domain structures can also be formed here, also resulting from the competing effects of the positive and negative current transients, although these are far less pronounced in the same fluence and pulse width range.

Further increasing the fluence results in more multi-domain structures, as shown in Figure S18. In particular, larger fluences result in partial switching of the RL layer. Larger fluences result in increased temperatures and demagnetization of the RL, more pronounced at the top surface (see Figure S11 and Figure S12). This can result in thermal nucleation of reversed domains in the RL. Moreover, spin torques are also generated on the RL due to the FL, in a similar way. Transfer of angular momentum from the FL to the RL is obtained during the positive current transient. This causes switching from AP+ to P- in the RL. During the negative current transient accumulation of minority spins at the FL causes switching from P+ to AP- in the RL. For larger fluences these three effects result in multi-domain structures in both the RL and FL, and for both starting states as shown in Figure S18.

## Switching probability maps and threshold fluences

The results plotted here show further details to the switching probability maps discussed in the main text, Figure 2. In particular, the low and high superdiffusive regime switching probabilities are re-plotted in Figure S20(a) and Figure S21(a) respectively, with low spin pumping regime, now also showing the high spin pumping regime for each case, Figure S20(b) and Figure S21(b) respectively. The extracted threshold fluences for P+ to AP+ and AP+ to P+ switching, as a function of pulse width, are shown in Figure S22 for the high superdiffusive regime – the corresponding threshold fluences for the low superdiffusive regime were discussed with reference to Figure 3 in the main text. Similar conclusions hold here, namely the superdiffusive mechanism dominates both the P+ to AP+, and AP+ to P+ switching. Moreover, no difference is observed for P+ to AP+ switching between the two spin pumping regimes, owing to the stronger superdiffusive regime used here.

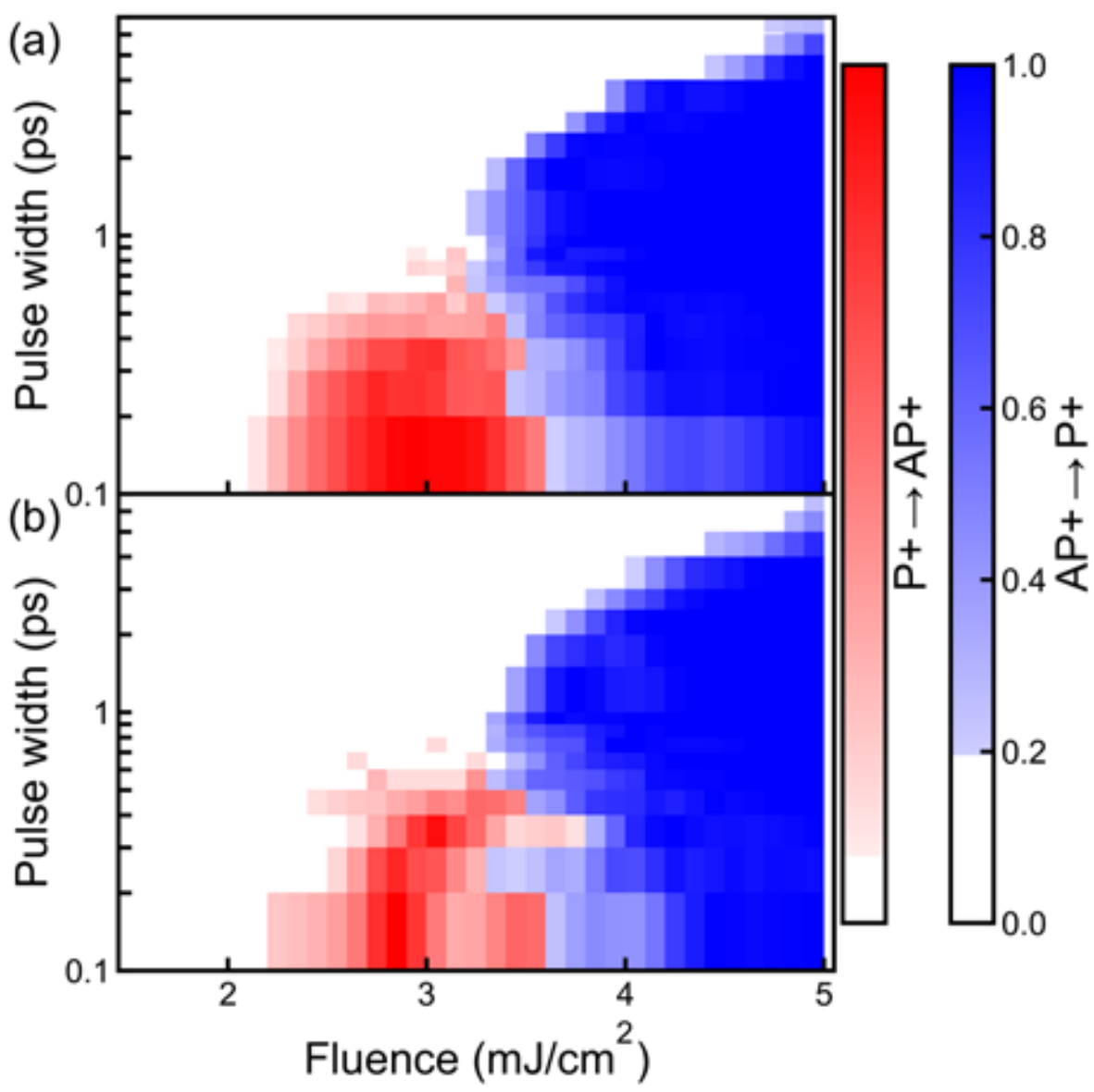

Figure S20 – Switching probability as a function of a pulse width and laser fluence for the low superdiffusive regime with $v_e$ = 0.25×10$^6$ m/s. The red and blue scales indicate P+ to AP+, respectively AP+ to P+, switching probability. (a) Low spin pumping regime with $\chi_{sp}$ = 2.4 kA/m, and (b) high spin pumping regime with $\chi_{sp}$ = 24 kA/m.

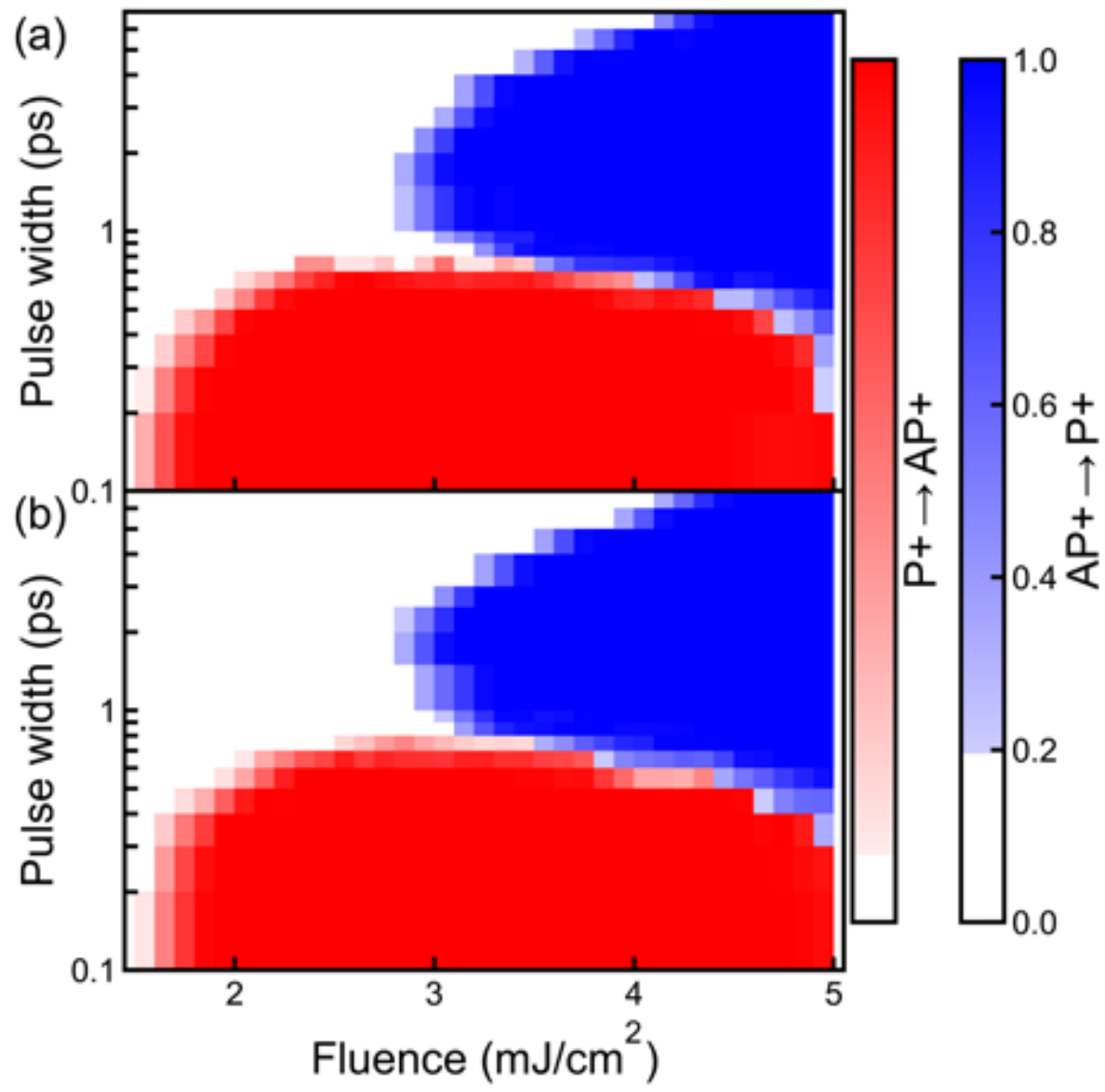

Figure S21 – Switching probability as a function of a pulse width and laser fluence for the high superdiffusive regime with $v_e$ = 1.00×10$^6$ m/s. The red and blue scales indicate P+ to AP+, respectively AP+ to P+, switching probability. (a) Low spin pumping regime with $\chi_{sp}$ = 2.4 kA/m, and (b) high spin pumping regime with $\chi_{sp}$ = 24 kA/m.

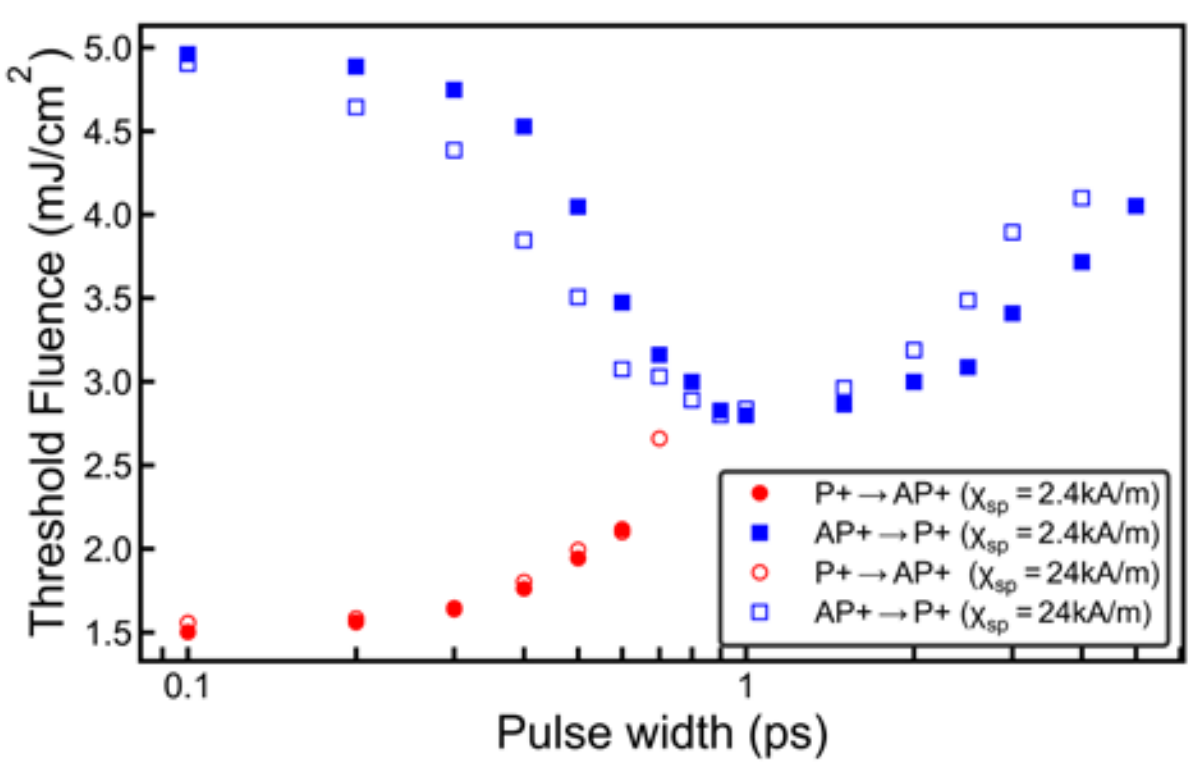

Figure S22 – Threshold fluence as a function of pulse width for P+ to AP+ (circles), and AP+ to P+ switching (squares), respectively. The threshold values are obtained from the high superdiffusive regime results in Figure S21, shown for both low (closed symbols) and high (open symbols) spin pumping regimes.